\documentclass{aa}  

\usepackage{graphicx}
\usepackage{txfonts}

\usepackage{natbib}      

\usepackage{soul}
\usepackage[dvipsnames]{xcolor}
\definecolor{DarkMagenta}{RGB}{180,0,139}  
\usepackage[colorlinks=true, linkcolor=DarkMagenta, citecolor=DarkMagenta, urlcolor=DarkMagenta]{hyperref}
\usepackage{soul}
\usepackage{subcaption}
\usepackage{ulem}
\usepackage{cancel}
\usepackage{graphicx}	
\usepackage{amsmath}	
\usepackage{mathtools,amssymb}
\usepackage{comment}

\newcommand{\micron}{$\mu$m}

\newcommand{\tdust}{$T_{\rm dust}$~}
\newcommand{\betadust}{$\beta_{\rm dust}$~}

\definecolor{checazneso}{rgb}{0.8, 0.8, 0.0}

\definecolor{amethyst}{rgb}{0.8, 0.0, 0.0}

\definecolor{darkgreen}{RGB}{0,200,0}

\usepackage{txfonts}
\begin{document}

   \title{The Drivers of Cosmic Dust Temperature Evolution}

   \subtitle{}

   \author{M. Parente
          \inst{1, 2}\fnmsep\thanks{\email{parente.m@ufl.edu}}
          \and
          F. Salvestrini\inst{1, 3}\fnmsep\thanks{\email{francesco.salvestrini@inaf.it}; second author with equal contribution.}
          \and
          G. L. Granato\inst{1,3,4} 
          \and  
          D. Narayanan\inst{2,5} 
          \and \\
          R. Tripodi\inst{6, 7, 3}
          \and
          S. Bianchi\inst{8}
          \and
          M. Bischetti\inst{1, 3, 9}
          \and
          C. Feruglio\inst{1, 3}
          \and
          F. Fiore\inst{1, 3}
          \and
          L. Silva\inst{1,3}
          }

    \institute
   {INAF, Osservatorio Astronomico di Trieste, via Tiepolo 11, I-34131, Trieste, Italy
    \and
    Department of Astronomy, University of Florida, 211 Bryant Space Sciences Center, Gainesville, FL 32611, USA 
    \and 
    IFPU, Institute for Fundamental Physics of the Universe, Via Beirut 2, 34014 Trieste, Italy
    \and
    IATE - Instituto de Astronom\'ia Te\'orica y Experimental, Consejo Nacional de Investigaciones Cient\'ificas y T\'ecnicas de la\\ Rep\'ublica Argentina (CONICET), Universidad Nacional de C\'ordoba, Laprida 854, X5000BGR, C\'ordoba, Argentina
    \and
    Cosmic Dawn Center at the Niels Bohr Institute, University of Copenhagen and DTU-Space, Technical University of Denmark
    \and
    INAF - Osservatorio Astronomico di Roma, Via Frascati 33, I-00078 Monte Porzio Catone, Italy
    \and
    University of Ljubljana FMF, Jadranska 19, 1000 Ljubljana, Slovenia
    \and
    INAF–Osservatorio Astrofisico di Arcetri, Largo E. Fermi 5, I-50125 Florence, Italy
    \and
    Dipartimento di Fisica “Enrico Fermi”, Università di Pisa, Largo Bruno Pontecorvo 3, Pisa I-56127, Italy
    }

   \date{Accepted July 10, 2026}

  \abstract
   {Observations of the rest-frame far-infrared (far-IR) emission of galaxies suggest a mild increase of dust temperature \tdust with redshift, although constraining \tdust in high-redshift systems remains challenging due to limited sampling of the far-IR spectral energy distribution (SED).}
   {We present and discuss the redshift evolution of \tdust predicted by a cosmological galaxy evolution simulation with dust treatment, and interpret its dependence on other galaxy physical properties.}
   {We use a semi-analytic model of galaxy formation that includes an explicit treatment of dust, post-processed with radiative transfer. Dust temperatures are derived by applying modified blackbody SED fitting to the simulated galaxies, mirroring the methodology adopted in most observational studies.}
   {The dust temperature of simulated galaxies increases with redshift, in broad agreement with observational results. A feature-importance analysis reveals that the star formation rate surface density $\Sigma_{\rm SFR}$ and the dust-to-gas ratio (DTG) are the main drivers of dust temperature, tracing the intensity of the interstellar radiation field and the optical depth of warm molecular clouds, respectively. Galaxies with higher star formation rate surface density and lower DTGs -- common conditions at high$-z$ -- are associated with warmer dust. We provide a simple relation to estimate DTG from $\Sigma_{\rm SFR}$, $T_{\rm dust}$, and redshift. Variations in dust grain size and chemical composition have a negligible impact on $T_{\rm dust}$. Our results are particularly relevant to the study of dust properties with observations of high-z galaxies, where far-IR dust emission is not fully sampled.}
   {}


   \maketitle

%

\section{Introduction}

Among the various components that contribute to the baryonic mass of the Universe, dust represents an almost negligible fraction of the interstellar medium (ISM; e.g., \citealt{Galliano18}), specifically less than 1\%. Nevertheless, it plays a fundamental role in a wide range of physical processes: it acts as a catalyst for molecular hydrogen formation \citep[e.g.,][]{Wakelam2017}, depletes metals from the gas phase of the ISM (e.g., \citealt{Jenkins09}), and absorbs ultraviolet (UV) photons by stars and re-emits them in the infrared (IR) band \citep[e.g.][]{DraineLee84, Silva98, Narayanan18}. 
Indeed, dust emission has been detected from the IR to radio wavelengths in galaxies out to very high redshift (e.g., \citealt{Watson15, Bowler18, Hashimoto19, Tamura19, Inami22, Bakx25}) thanks to sensitive observations with both ground-based facilities such as the Atacama Large Millimeter Array (ALMA) or the Northern Extended Millimeter Array (NOEMA), and space telescopes such as \textit{Herschel}. As a result, dust has become a key tracer for inferring the physical properties of galaxies, such as star formation activity, gas content, and metal enrichment, and therefore plays a central role in studies of galaxy evolution \citep{Santini14, Genzel15, Scoville16, Algera23, Traina24a, Traina24b, Sun25}.\\
Two of the key dust properties that observations aim to constrain are the dust mass and temperature. The latter provides insight into fundamental physical properties of galaxies -- such as the intensity of the radiation field and the spatial distribution of stars relative to dust \citep{Silva98}. However, determining the temperature of dust grains within a galaxy is a challenging task, as it requires observations over a wide range of wavelengths, from a few tens of microns up to a few millimeters. Furthermore, determining the temperature is degenerate with the dust mass and the dust emissivity index, since the total IR luminosity scales as $L_{\rm IR} \propto M_{\rm dust} T_{\rm dust}^{4+\beta_{\rm dust}}$, where $\beta_{\rm dust}$ is the emissivity index (e.g., \citealt{Ferrara22}). Accurately determining the dust temperature is therefore crucial both for gaining physical understanding of the conditions of the ISM and for deriving robust estimates of the dust mass.\\
Over the decades, studies of dust temperature have been conducted for both star-forming galaxies and AGN, generally revealing similar trends and evolutionary behaviors \citep[e.g.,][]{Witstok23, Tripodi24b}.
In many cases, however, the two populations are treated similarly, even though AGN are expected to contribute directly to dust heating even at kiloparsec scales \citep[e.g.,][]{Tsukui23, FernandezAranda25} although the importance of AGN-heated dust remains debated, and in some cases labeled as   subdominant \citep[e.g.,][]{Silverman26}. This reflects the intrinsic complexity of measuring dust temperatures, which requires broad multi-wavelength coverage and whose reliability strongly depends on the availability and spectral sampling of the photometric data. Some studies have adopted multi-temperature fitting functions and composite models \citep[e.g.,][]{Dunne01, Casey12, Dale12, Kirkpatrick15}, while others developed empirical templates of the spectral energy distribution (SED) based on distributions of interstellar radiation field intensities heating the dust grains \citep[e.g.,][]{DraineLi07, Galliano18b}. However, these approaches are characterized by a large number of free parameters and are therefore difficult to apply to large samples of galaxies, especially at high redshift ($z>2$), where only sparse dust continuum detections are often available. In such cases, fitting the dust SED with a single-temperature modified blackbody (MBB) represents a reasonable and widely adopted approximation \citep[e.g.,][]{Hildebrand83, Magnelli12, Simpson17, Carniani19, Bakx21, Witstok23, Tripodi24b}.\\
Regardless of the method employed or the quality of the SED sampling, it is important to recognize that most observational analyses assume a single, isothermal dust temperature. In reality, galaxies contain broad, multi-temperature dust distributions arising from the combination of diffuse ISM and compact star-forming regions \citep{Draine03}. Consequently, the commonly adopted single-temperature approximation can bias dust mass estimates \citep{Sommovigo&Algera25}. Dust temperature measurements obtained in this way should therefore be regarded as effective, model-dependent parameters rather than direct physical temperatures.\\
Taken together, current observational studies suggest that the average effective dust temperature of star-forming galaxies increases with redshift. Measurements spanning from the local Universe to early cosmic times -- based on both individual detections and stacked samples -- point to a systematic, albeit modest, rise in dust temperature (e.g., \citealt{Schreiber18, Viero22, Witstok23}), while the inferred dust temperature distribution at redshift $z>1$ remains extremely broad. However, observations at high $z$ are subject to strong selection effects, as only the most massive, intensely star-forming, and dust-rich systems are typically detected \citep[e.g.,][]{Lim20}. Consequently, the detailed form of the temperature–redshift relation is still poorly constrained and depends sensitively on sample selection, analysis techniques, and assumptions regarding dust physics.\\
From a theoretical perspective, numerous studies have aimed to predict dust temperatures in galaxies using a variety of approaches. The most widely adopted and physically motivated method involves post-processing galaxy formation simulations with radiative transfer (RT) codes (e.g., \textsc{SKIRT}; \citealt{SKIRT}) to model dust heating, temperature distributions, and the resulting IR emission \citep[e.g.][]{Behrens18, Aoyama19, Liang19, Ma2019, Pallottini22, Vijayan22, Shen22}. While these studies differ substantially in their modeling strategies and in how dust temperature is defined, they broadly agree on a mild increase of dust temperature with redshift, primarily driven by higher specific star formation rates (SFRs) and SFR surface densities. It is nevertheless important to note that, despite growing efforts within the numerical extragalactic community to explicitly model dust in galaxy evolution simulations (e.g., \citealt{Parente25rev}), the majority of the works above do not include such treatments and instead assume a fixed dust-to-metal ratio to link dust to the metal content \citep[but see][]{Granato21, Lower24}.\\
In parallel, a number of simple yet physically motivated analytic models have been developed to address this problem.
For instance, \cite{Sommovigo22a} associates the higher dust temperatures observed at high redshift to shorter gas depletion times driven by more intense cosmological accretion at early epochs, while \cite{Hirashita22} attributes elevated \tdust at high $z$ to the combined effects of low dust-to-gas (DTG) ratios and enhanced star formation activity.\\

In this work, we build a theoretical framework to investigate the evolution of dust temperature with redshift in star-forming galaxies by combining all the key ingredients required for this purpose. Specifically, we employ: \textit{(i)} a cosmological galaxy evolution simulation with explicit dust modeling \citep{Parente2023}; \textit{(ii)} explicit RT calculations to model dust emission \citep{Silva98}; and \textit{(iii)} an observationally motivated dust temperature determination obtained by fitting the SEDs derived from our simulations \citep{Tripodi24b}. We present predictions for the redshift evolution of $T_{\rm dust}$, which we compare against an extensive compilation of observational measurements in several samples of galaxies and AGN from the local Universe up to $z\sim 8$, and we further explore its dependence on key galaxy properties, most notably the SFR surface density and the DTG ratio.\\
The paper is organized as follows: the results of the simulation used in this work are presented in Sect.~\ref{sec:sim_cat}, including our analysis to derive the estimates of $T_{\rm dust}$. Sect.~\ref{sec:results} presents the analysis results, compares them with literature findings, and provides a physical interpretation of our modeled \tdust and its dependence on key galaxy properties. Finally, the results are summarized in Sect.~\ref{sec:conclusions}.
\\

\section{Simulation catalog}
\label{sec:sim_cat}
Our catalog of simulated galaxies is built using the \textsc{L-Galaxies} semi-analytic model (SAM), adopting its latest public release \citep{Henriques2020}\footnote{The source code is available at \url{https://github.com/LGalaxiesPublicRelease/LGalaxies_PublicRepository/releases/tag/Henriques2020}}
 together with the updates presented in \cite{Parente2023}. Among these updates, particularly relevant for this work is the implementation of a new model for the formation and evolution of dust grains, inspired by the dust prescriptions previously developed for hydrodynamical simulations \citep{Granato21, Parente22}.

\begin{figure}
    \centering
    \includegraphics[width=0.99\columnwidth]{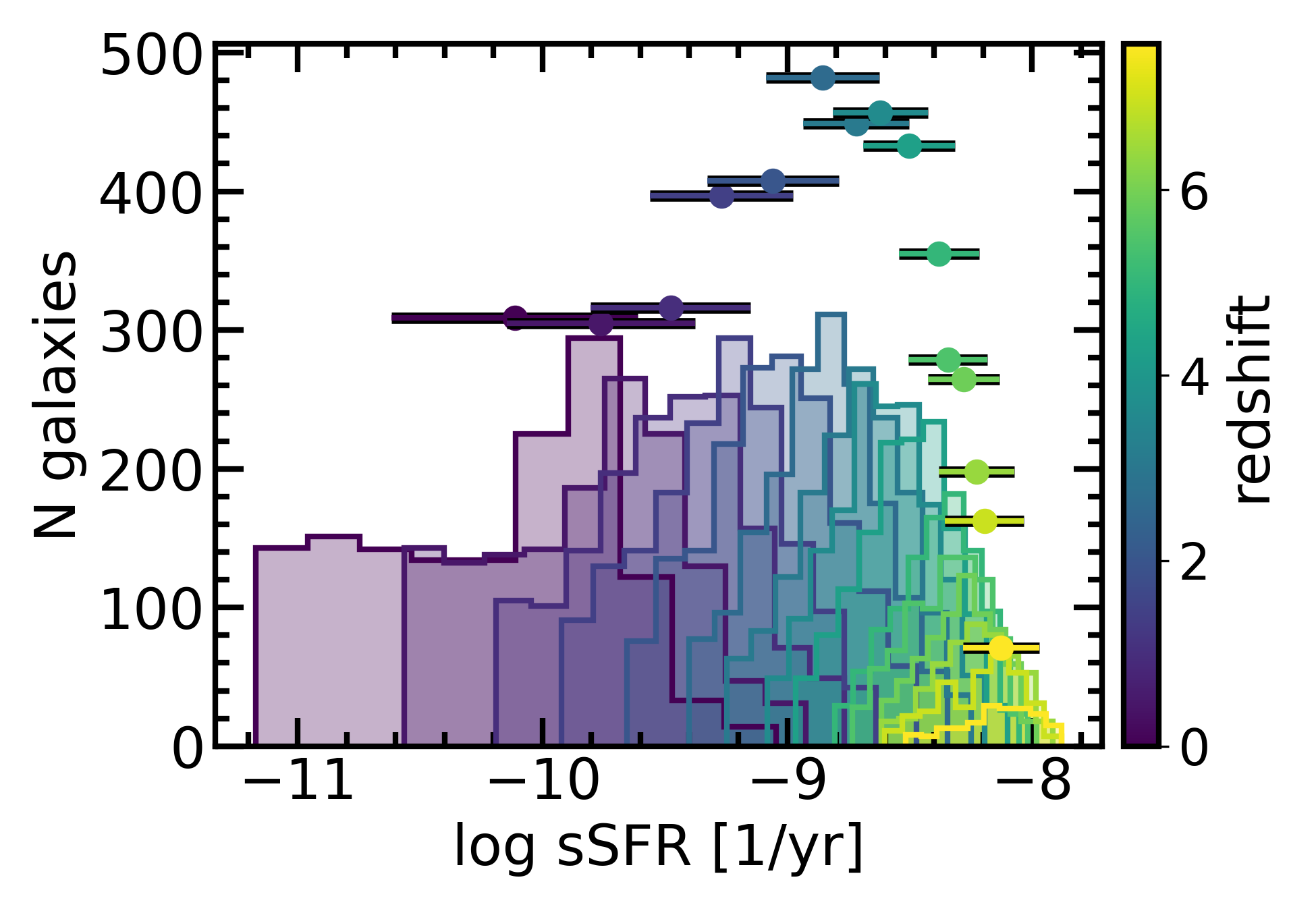}
    \caption{Distribution of the specific SFR of the simulated galaxies used in the analysis across different redshifts (see colorbar). Circles and horizontal bars indicate the median and standard deviation of the distribution at each redshift.}
    \label{fig:sSFRhisto}
\end{figure}

The dust model (for a detailed description, see Sect. 2.1 of \citealt{Parente2023}) includes two grain sizes (small and large, with radii of $0.005$ and $0.05\,\mu$m respectively) and two chemical compositions (carbonaceous and silicate). Dust grains are produced in the envelopes of AGB stars and in the ejecta of core-collapse supernovae, and are subsequently released into the ISM. Their mass and size evolution is regulated by several processes, including accretion, destruction in supernova shocks, thermal sputtering, shattering, and coagulation. The resulting dust masses and grain properties are a key ingredient for predicting the SED of galaxies.\\
The SAM is run on the \textsc{Millennium} merger trees (\citealt{Springel05}; $500\,{\rm Mpc}/h$ box size), rescaled to a \textit{Planck} cosmology\footnote{The original \textsc{Millennium} cosmology is rescaled following \cite{Angulo2010,Angulo2015}.} \citep{Planck14} with $h=0.673$, $\Omega_{\rm m}=0.315$, $\Omega_{\rm b}=0.0487$, and $\sigma_8=0.829$. A \cite{Chabrier03} IMF is adopted. In our analysis we consider galaxies with $\log(M_{\rm stars}/M_\odot) \geq 9$, close to the resolution limit of the underlying dark matter simulation \citep[e.g.][]{Guo11}. Passive galaxies are excluded, defined as those with specific SFR (sSFR) more than $1.5\sigma$ below the median\footnote{The median is calculated using galaxies with $M_{\rm stars} < 10^{10}\,M_\odot$ to avoid bias from massive quiescent systems.} at each redshift.
To reduce computational cost, we post-process a representative subsample of galaxies per redshift with radiative transfer, uniformly sampling the full stellar-mass range. This results in $\approx 2,500$ galaxies at $z=0.0$ and $\approx 250$ galaxies at $z=7.57$. The sSFR distribution of our simulated galaxy sample is shown in Fig. \ref{fig:sSFRhisto}.

\subsection{Simulated SEDs with radiative transfer}
\label{sec:grasil}


In order to obtain the SED of our simulated galaxies, we exploit the radiative transfer code \textsc{GRASIL} \citep{Silva98,Granato00}, which models stellar emission and its interaction with dust, including absorption and thermal re-emission. GRASIL uses the star formation histories, galaxy geometry, and dust masses and properties as predicted by the SAM. While we briefly summarize its strategy in the following, we refer to \cite{Parente2025GV} for a description of the coupling between this RT pipeline and the SAM.\\
Galaxies are modeled with a disk-like ISM and a stellar component composed of a disk and a bulge. Stellar and ISM disks follow exponential radial and vertical profiles, with scale lengths provided by the SAM, while the bulge follows a King profile. Stellar emission is computed by convolving the disk and bulge star formation histories with the simple stellar population models by \cite{bressan1998, bressan2002}, based on the Padova stellar isochrones \citep{Bertelli94, Bressan94} covering ages of $1\,{\rm Myr}-20\,{\rm Gyr}$ and metallicities $Z=0.0004-0.05$.\\
RT accounts for the interaction between starlight and dust grains, whose abundances, compositions, and sizes are predicted by the SAM. Dust is distributed between molecular clouds and a diffuse ISM, with the molecular fraction provided by the SAM. Stars are assumed to form in molecular clouds and escape on a characteristic timescale $t_{\rm esc}=3\, {\rm Myr}$, after which their radiation propagates in the diffuse medium. Full RT calculations are performed within molecular clouds, while a simplified treatment is adopted for the diffuse phase.
The resulting stellar and dust emission is combined to produce galaxy SEDs spanning $0.1-10^4\,\mu$m.

\subsection{SED binning}
\label{sec:sedbin}
Given the SED obtained from the RT post-processing described in Sect.~\ref{sec:grasil} for galaxies spanning the redshift range $0 \lesssim z \lesssim 7.5$, we extracted flux densities from the model SED over the rest-frame wavelength interval 40~\micron--3~mm.
To mimic realistic observations, we selected the flux corresponding to the wavelength closest to the effective wavelength of representative bands in the far-IR regime.
Specifically, we considered the 100 and 160~\micron\ bands from the Photodetector Array Camera and Spectrometer (PACS) and the 250~\micron\ band from the Spectral and Photometric Imaging Receiver (SPIRE), both onboard the \textit{Herschel} satellite. In addition, we included ALMA bands covering observed-frame wavelengths from 3.5~mm to 0.35~mm, corresponding to Bands~2 and~10\footnote{\url{https://www.eso.org/public/italy/teles-instr/alma/receiver-bands/}}, respectively.
We excluded the \textit{Herschel}/SPIRE 350 and 500~\micron\ bands, as they overlap with ALMA Bands~10 and~9, respectively.
This procedure yielded a set of 8 or 9 photometric points for each SED, depending on the galaxy redshift, sampling both sides of the peak of the far-IR emission as well as the Rayleigh–Jeans regime.
The full list of representative wavelengths adopted for each redshift bin is provided in Table \ref{tab:bands}.

We acknowledge that assuming full spectral coverage is overly optimistic, as this condition is rarely met in real life. While this choice allows us to focus on the intrinsic predictions of the model and to maximize the robustness of the link between \tdust and galaxy physical properties, we note that sparse photometric coverage is often a major source of uncertainty in dust temperature estimates \citep[e.g.,][]{Casey12}.

\begin{table}[ht]
\centering
\caption{Photometric bands adopted in the SED fitting.}
\label{tab:bands}         
\begin{tabular}[0.5\textwidth]{cl}     
\hline                        
$z$ & Bands \\    
\hline 
0.05	& 	PACS100, PACS160, SPIRE250, 10, 9, 8, 7, 6, 5   \\
0.51	& 	PACS160, SPIRE250, 10, 9, 8, 7, 6, 5, 4  \\
1.04	& 	PACS160, SPIRE250, 10, 9, 8, 7, 6, 5, 4   \\
1.48	& 	PACS160, SPIRE250, 10, 9, 8, 7, 6, 5, 4   \\
2.07	& 	SPIRE250, 10, 9, 8, 7, 6, 5, 4, 3   \\
2.64	& 	SPIRE250, 10, 9, 8, 7, 6, 5, 4, 3   \\
3.11	& 	SPIRE250, 10, 9, 8, 7, 6, 5, 4, 3   \\
3.65	& 	SPIRE250, 10, 9, 8, 7, 6, 5, 4, 3   \\
4.28	& 	10, 9, 8, 7, 6, 5, 4, 3   \\
5.03	& 	10, 9, 8, 7, 6, 5, 4, 3   \\
5.46	& 	10, 9, 8, 7, 6, 5, 4, 3   \\
5.92	& 	10, 9, 8, 7, 6, 5, 4, 3   \\
6.42	& 	10, 9, 8, 7, 6, 5, 4, 3, 2   \\
6.97	& 	10, 9, 8, 7, 6, 5, 4, 3, 2   \\
7.57	& 	10, 9, 8, 7, 6, 5, 4, 3, 2   \\
\hline
\end{tabular}
\flushleft 
\caption*{\small Note: Redshift bins and the corresponding photometric bands used to bin each galaxy SED, listed in increasing wavelength order. PACS and SPIRE stand for \textit{Herschel} PACS and SPIRE instruments; the numbers between 2 and 10 indicate the corresponding ALMA bands.}
\end{table}

\subsection{Fit of the far-IR SED}
\label{sec:dustsed}


\begin{figure}
    \centering
    \includegraphics[width=0.9\columnwidth]{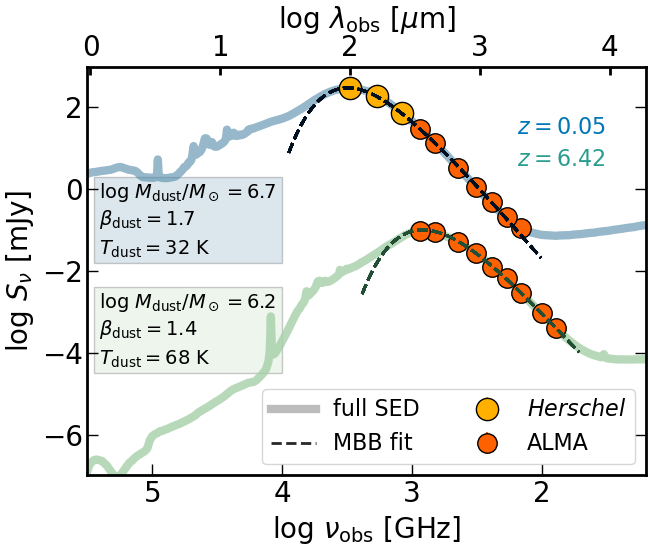}
    \caption{Example of the SED fitting procedure for two model galaxies at $z=0.05$ (blue) and $z=6.42$ (green). Frequencies and wavelengths are in the observer frame. The full SED from our model is shown as a solid line, while the binned points -- representing mock \textit{Herschel} and ALMA observations -- are plotted as symbols. The dashed lines show the resulting MBB fit to these points, with the fitted parameters of both SEDs reported in the top-left corner.}
    \label{fig:sed}
\end{figure}

To fit the mock photometric points of the SEDs we used the publicly available {\sc EOS-Dustfit} code\footnote{{\sc EOS-Dustfit} is a publicly available tool for fitting the cold dust SED of galaxies (\url{https://github.com/roberta96/EOS-Dustfit}).
Details about the modeling can be found on the GitHub page and in \citet{Tripodi24b, Salvestrini25}.}.
In {\sc EOS-Dustfit}, the cold dust continuum emission is modeled with a single-temperature modified blackbody (MBB) function, assuming an optically thick regime, given by:
\begin{equation}
\label{eq:Snufit}
S_{\nu_{\rm obs}}^{\rm obs} = 
\frac{\Omega}{(1+z)^3}
\left[ B_\nu\!\left(T_{\rm dust}(z)\right)
      - B_\nu\!\left(T_{\rm CMB}(z)\right) \right]
\left(1 - e^{-\tau_\nu}\right),
\end{equation}
where $\Omega = (1+z)^4 A_{\rm gal} D_{\rm L}^{-2}$ is the solid angle, and
$A_{\rm gal}$ and $D_{\rm L}$ are the projected surface area and the luminosity
distance of the galaxy, respectively. The dust optical depth is defined as
\begin{equation}
\label{eq:tau}
\tau_\nu =
\frac{M_{\rm dust}}{A_{\rm gal}}
\, k_0 \left( \frac{\nu}{250\,{\rm GHz}} \right)^{\beta_{\rm dust}},
\end{equation}
where $\beta_{\rm dust}$ is the dust emissivity index and
$k_0 = 0.45\,{\rm cm}^2\,{\rm g}^{-1}$ is the mass absorption coefficient \citep{Beelen06}. The impact of our assumption of optically thick medium is minor, and briefly discussed in App. \ref{app:opt_thick}.\\
The effect of the cosmic microwave background (CMB) on the dust temperature is
accounted for through
\begin{equation}
T_{\rm dust}(z) =
\left[
(T_{\rm dust})^{4+\beta_{\rm dust}}
+ T_0^{\,4+\beta_{\rm dust}}\left( (1+z)^{4+\beta_{\rm dust}} - 1 \right)
\right]^{1/(4+\beta_{\rm dust})},
\end{equation}
where $T_0 = 2.73\,{\rm K}$. 
The model also includes the contribution of the CMB emission, described by the Planck function
$B_\nu\!\left(T_{\rm CMB}(z)\right) = B_\nu\!\left(T_0 (1+z)\right)$ \citep{daCunha13}.\\
The solid angle $\Omega$ is estimated assuming that the size of the emitting region is 1.5 times the scale radius obtained from fitting the radial profile of the cold gas with an exponential function\footnote{We tested how sensitive our results are to the assumed source size by doubling and halving the radius. The three best-fitting parameters remain consistent with our baseline results within the uncertainties. For instance, doubling the size lowers the median $T_{\rm dust}$ by at most $\approx 2$ K.}. This region encompasses nearly 90\% of the dust in the simulated galaxy, which should roughly match the bulk of the dust-continuum emission that is usually collected with submillimeter observations, especially at high redshift \citep[e.g.,][]{Hodge16, Gullberg19}.
In our analysis, we left three parameters in the MBB model free to vary, namely dust temperature ($T_{\rm dust}$), dust mass ($M_{\rm dust}$), and the emissivity index ($\beta_{\rm dust}$).
A conservative 10\% uncertainty was adopted to account for calibration inaccuracies inherent to observations.
{\sc EOS-Dustfit} explores the parameter space for each SED using a Markov chain Monte Carlo algorithm implemented in the \textit{emcee} package \citep{emcee1}. 
A uniform distribution for priors is assumed for fitting parameters in the range $5\leq T_{\rm dust}/{\rm K}\leq300$, $3\leq\log(M_{\rm dust}/M_{\odot})\leq9$, and $0.5\leq\beta_{\rm dust}\leq3.0$. 
We ran 30 chains, each with 500 trials and a burn-in phase of 200 steps, for a total of 15,000 steps per galaxy. 
We adopted the 50th percentile of the posterior distribution as the best-fit value, while the errors are calculated considering the 16th and 84th percentiles.
The results of the SED fitting are presented in Sect.~\ref{sec:results}, while Fig. \ref{fig:sed} provides an explicit example of the procedure for two model galaxies at $z=0.05$ and $z=6.42$.\\

\section{Results}
\label{sec:results}

\subsection{Redshift evolution of \tdust}
\label{sec:tdust_evol_z}


\begin{figure*}
    \centering
    \includegraphics[width=1.75\columnwidth]{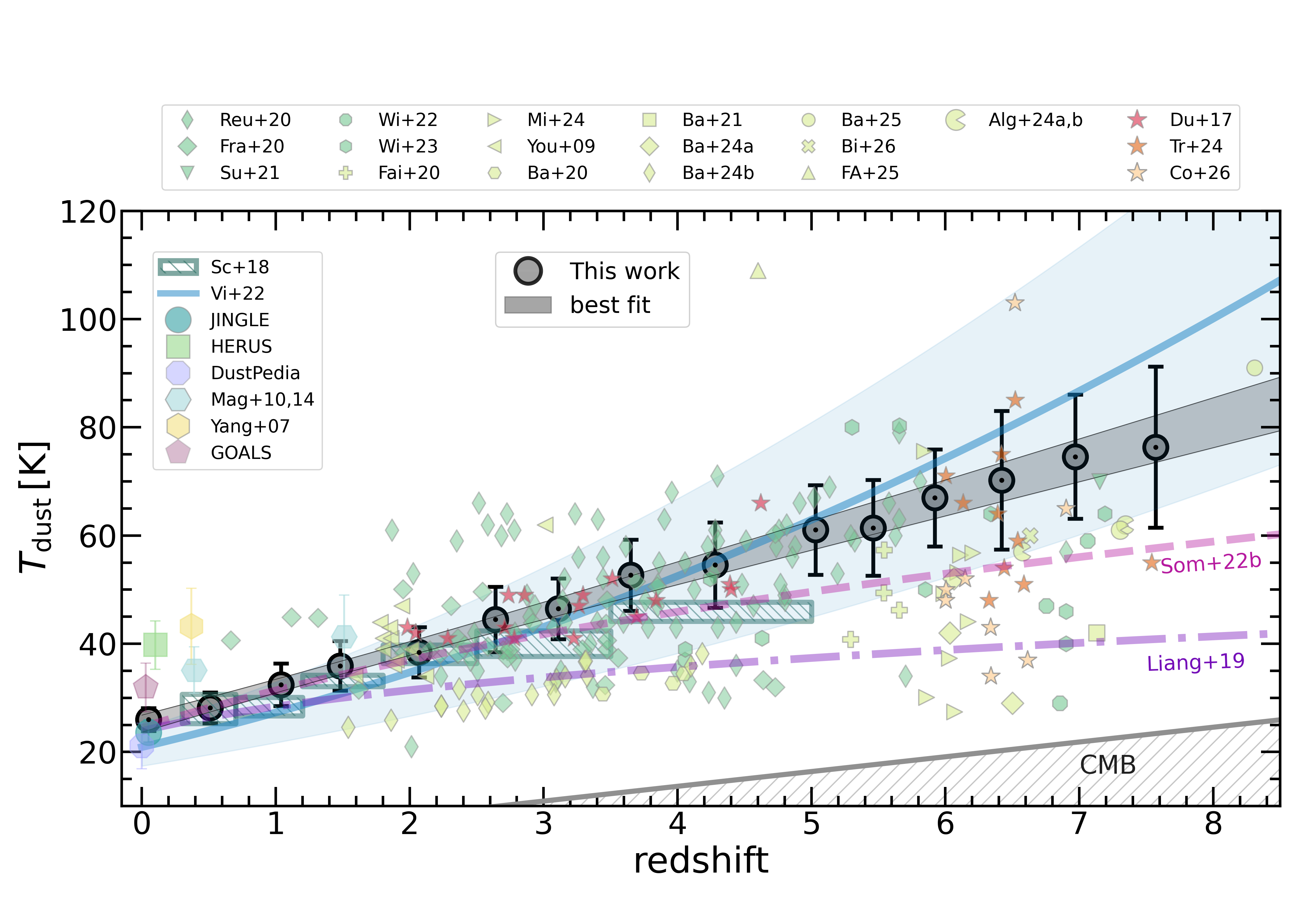}
    \caption{Dust temperature evolution with redshift. Our results from simulated galaxies are shown as gray circles (median values), with error bars indicating the $16$–$84$th percentile dispersion. A linear fit to our results is displayed as a shaded gray region (Eq. \ref{eq:Tzfit}).\\ 
    Measurements from the literature are also included for comparison (see extensive description in App. \ref{app:obs_data}).
    Owing to the relatively limited redshift range of some samples, which would make the distribution of individual objects difficult to visualize, we report the median $T_{\rm dust}$ values and their standard deviations (HERUS from \citealt{Clements18}; JINGLE from  \citealt{Lamperti19}; DustPedia from \citealt{Nersesian19}; \citealt{Magdis10, Magdis14}; \citealt{Paspaliaris21}; \citealt{Yang07}). Large samples extending to $z > 2$ are those presented by \cite{Schreiber18} and \cite{Viero22} (Sc+18 and Vi+22, respectively, both obtained from stacking analyses.).
    Less populated datasets -- shown as individual galaxies -- are also included: \cite{algera2024b, algera2024, Bakx20b, Bakx21, Bakx24, Bakx24b, Bakx25, Bing26, Faisst20, Franco20, Mitsuhashi24, Reuter20, Sugahara21, Witstok22b, Witstok23,  Younger09}.
    We also include observational samples of high-redshift AGN and QSOs \citep{Duras17, Bischetti21, Tripodi24b, Costa26, FernandezAranda25}. Theoretical predictions from the physical model of \cite{Sommovigo22a} and the numerical simulations of \cite{Liang19} are shown as shaded lines. For reference, the CMB temperature is indicated by a hatched region.}
    \label{fig:Tdust_z}
\end{figure*}

We present the redshift evolution of the dust temperature, $T_{\rm dust}$, in Fig. \ref{fig:Tdust_z}, comparing our results with a compilation of measurements from the literature. All data points and trends shown in the figure are based on observational studies, with the sole exception of the relation from \cite{Liang19} and \citet{Sommovigo22a}, which represent theoretical predictions.\\
For each redshift bin, we report the median value and $16-84$th dispersion of the \tdust measurements for the simulated galaxy population.
Our results exhibit a clear redshift evolution of $T_{\rm dust}$, increasing from $\approx 20 \, {\rm K}$ in the local Universe up to $\approx 70 \, {\rm K}$ at $z \approx 7.5$.
This trend is in good agreement with the behavior inferred from previous results in the literature.
Using the \texttt{polyfit}\footnote{\url{https://numpy.org/doc/stable/reference/generated/numpy.polyfit.html}} routine from the Python \texttt{numpy} package, we perform a linear fit, finding:
\begin{equation}
\label{eq:Tzfit}
    T_{\rm dust} = (6.95 \pm 0.68)\times z \, + \, (25.2 \pm 1.6).
\end{equation}

We note that the dispersion of the \tdust measurements increases with redshift, from $\approx 2 \, {\rm K}$ in the local Universe to $\approx 15 \, {\rm K}$ -- roughly $20\%$ of the temperature in the highest redshift bin. The increase of the \tdust dispersion with redshift is briefly discussed in light of the temperature dependence on SFR surface density and DTG in Sect.~\ref{sec:scalingscatter}.

\subsubsection{Comparison with observations}
\label{sec:comparison_data}

The results of our analysis are broadly consistent with the observational results from the literature\footnote{We remind that we compiled observational results obtained using methodologies broadly similar to those adopted in our analysis. Whenever possible, we verified the consistency among the different approaches, finding overall good agreement. For the sake of brevity, we limit the discussion here to a direct comparison with the results available in the literature. Additional details on the methodologies employed and on the specific assumptions adopted in each study are summarized in App. \ref{app:obs_data}. We also refer the reader to the original references cited throughout the text for a comprehensive description.}.\\
In the nearby Universe, our analysis predicts a \tdust which is close to that derived for the local star-forming galaxies (SFGs), as those analyzed in the JINGLE survey \citep{Lamperti19} and in DustPedia \citep{Nersesian19}.
On the contrary, the median value of \tdust measured in the HERUS sample \citep{Clements18}, which includes local ultra-luminous infrared galaxies (ULIRGs), is roughly 3$\sigma$ higher than our predictions.
A similar behavior is also observed for samples of ULIRGs at $z < 1$, such as those presented by \citet{Yang07}, \citet{Magdis14}, and \citealt{Paspaliaris21}. 
The relatively high \tdust value measured in ULIRGs when compared to normal SFGs is likely driven by the significantly larger amount of stellar emission absorbed by dust particles and the relatively higher contribution from young stellar population to the interstellar radiation field with respect to typical early and late-type galaxies in the local Universe (e.g., \citealt{Magdis10, Paspaliaris21, Hogan22})\\
A similar dichotomy between SFGs and ULIRGs is observed at intermediate redshifts ($1 \lesssim z \lesssim 4$), with our results closely following the empirical trends reported in the literature (\citealt{Schreiber18, Viero22}), and within the observed scatter defined by samples presented by \cite{Younger09} and \cite{Franco20}. Even luminous quasars (QSOs) with accurate treatment of the reprocessing of the AGN emission, such as the WISSH sample by \cite{Duras17}, strictly follow our prediction of the evolution of \tdust with redshift.\\
At higher redshift ($z \gtrsim 5$), the observational data display a large dispersion, reflecting both intrinsic diversity among galaxies and increasing uncertainties due to limited photometric coverage. In this regime, our simulated galaxies populate the upper part of the observed $T_{\rm dust}$ distribution and are consistent with measurements for luminous systems such as those in the ALPINE, REBELS (e.g., \citealt{algera2024b, algera2024, Algera25c, Faisst20}) samples, as well as the objects from \cite{Witstok23} and \cite{Tripodi24b}.
Regarding \cite{Tripodi24b}, the authors adopted a different parametrization for the evolution of \tdust with redshift (see their Eq.~6). Nevertheless, the trends derived from both their full galaxy sample and their QSO-only sample are consistent with ours within the respective uncertainties.\\
We remind the reader that, to make the different observational samples easier to compare, we chose literature works that adopted broadly similar methodologies to measure \tdust (see App. \ref{app:obs_data}). 
The heterogeneous galaxy selection criteria and the sparse sampling of the far-IR emission adopted in the studies shown in Fig.~\ref{fig:Tdust_z}, particularly at $z>1$, likely contribute significantly to the observed dispersion.
Indeed, the spread in \tdust reaches up to $\approx 20$~K at Cosmic Noon ($1.5<z<4.5$), even among galaxies within the same sample (e.g., \citealt{Reuter20}).
This scatter likely arises from a combination of large measurement uncertainties (with $\Delta T_{\rm dust} / T_{\rm dust}$ reaching up to $\approx 30$–$40\%$) and differences in sample selection.
Notably, when comparing with larger samples such as that presented in \cite{Viero22}, the observed dispersion is found to increase with redshift, in agreement with our results.\\
Finally, we note that the observational samples considered also include galaxies hosting AGN and QSOs, whose contributions are not modeled in our simulation. Furthermore, even if most of the ULIRGs included here were selected to be dominated by star formation activity (see \citealt{Magdis14}), it is difficult to rule out any potential contamination from dust-obscured AGN activity obscured by the total far-IR emission.
Therefore, determining the actual impact of AGN activity on dust heating, even on kiloparsec scales, is far from trivial. Growing evidence indicates that AGN can contribute up to $\approx 50\%$ of the total far-IR emission (e.g., \citealt{Tsukui23}), with a pronounced spatial gradient in the peak dust temperature (e.g., \citealt{FernandezAranda25}). However, the substantial scatter characterizing the \tdust distribution, including that of high-redshift AGNs shown in Fig.~\ref{fig:Tdust_z}, suggests that the intensity of the AGN-powered ionizing radiation field in the galaxy center is not the only driving mechanism contributing to dust heating (see the discussion in Sect.~\ref{sec:scalingscatter}).

\subsubsection{Comparison with theoretical models}
\label{sec:comparison_model}

For the comparison with theoretical predictions, we consider the models of \citet{Sommovigo22a} and \citet{Liang19}. The former is based on an analytic framework that links dust temperature to the cosmological evolution of galaxy properties, in particular to gas depletion times, which are shorter in high-redshift galaxies owing to enhanced cosmological accretion rates (as also confirmed by observations; e.g. \citealt{Tripodi24b, Salvestrini25}).
While this model predicts an overall increase of $T_{\rm dust}$ with redshift, in agreement with the above-mentioned observational results up to intermediate redshifts ($z\sim4$), it yields systematically lower temperatures than those inferred from our analysis and from some observational samples at $z \gtrsim 5$. However, we note that the \cite{Sommovigo22a} relation presented here is based on their solar metallicity model. The authors point out that variations in column density and metallicity can lead to temperature differences of approximately $\Delta T_{\rm dust} \approx 25\,{\rm K}$. For instance, their optically thick model with a metallicity of $0.3 \, Z_\odot$ -- associated with lower DTG ratios, as expected at high redshift -- yields $T_{\rm dust} \approx 60 \, {\rm K}$ at $z \approx 6$, in better agreement with our model.\\
The work\footnote{We report their predicted peak temperature, defined as the temperature inferred from the wavelength at which the far-IR emission reaches its maximum. A discussion of the different methods used to estimate dust temperature is presented in the original work, to which we refer the reader.} by \cite{Liang19} instead is based on the analysis of massive galaxies at $z \approx 2$–$6$ using hydrodynamic simulations from the FIRE project \citep{FIRE}, coupled with RT calculations performed with \textsc{SKIRT} \citep{SKIRT}.
Although their results also show an increase in dust temperature with redshift, the inferred temperatures are systematically lower than ours, with the difference becoming larger at high redshift.
While a detailed comparison between their simulations and our semi-analytic framework is beyond the scope of this work, we here highlight two key differences that are likely to drive the aforementioned difference in $T_{\rm dust}$.\\
First, unlike our model, the \cite{Liang19} FIRE simulations do not include an explicit treatment of dust formation and evolution. Dust is therefore assumed to directly trace metals through a fixed dust-to-metals ratio of $0.4$. This assumption represents a significant source of uncertainty, particularly at high redshift, where our model predicts dust-to-metals ratios as low as $\approx 10^{-2}$. 
Second, the FIRE simulations achieve exceptionally high spatial resolution, allowing them to resolve the structure of the ISM. On the contrary, our SAM adopts a simplified geometric description of the ISM, which is modeled as a diffuse disk with embedded spherical molecular clouds. Given the crucial role of geometry in radiative transfer calculations \citep{Silva98} and its importance in shaping our results (see the discussion in Sect.~\ref{sec:shap}), this difference is likely another major contributor to the discrepancies.

\subsection{The driver of the \tdust$-z$ relation}
\label{sec:shap}


\begin{figure}
    \centering
    \includegraphics[width=1\columnwidth]{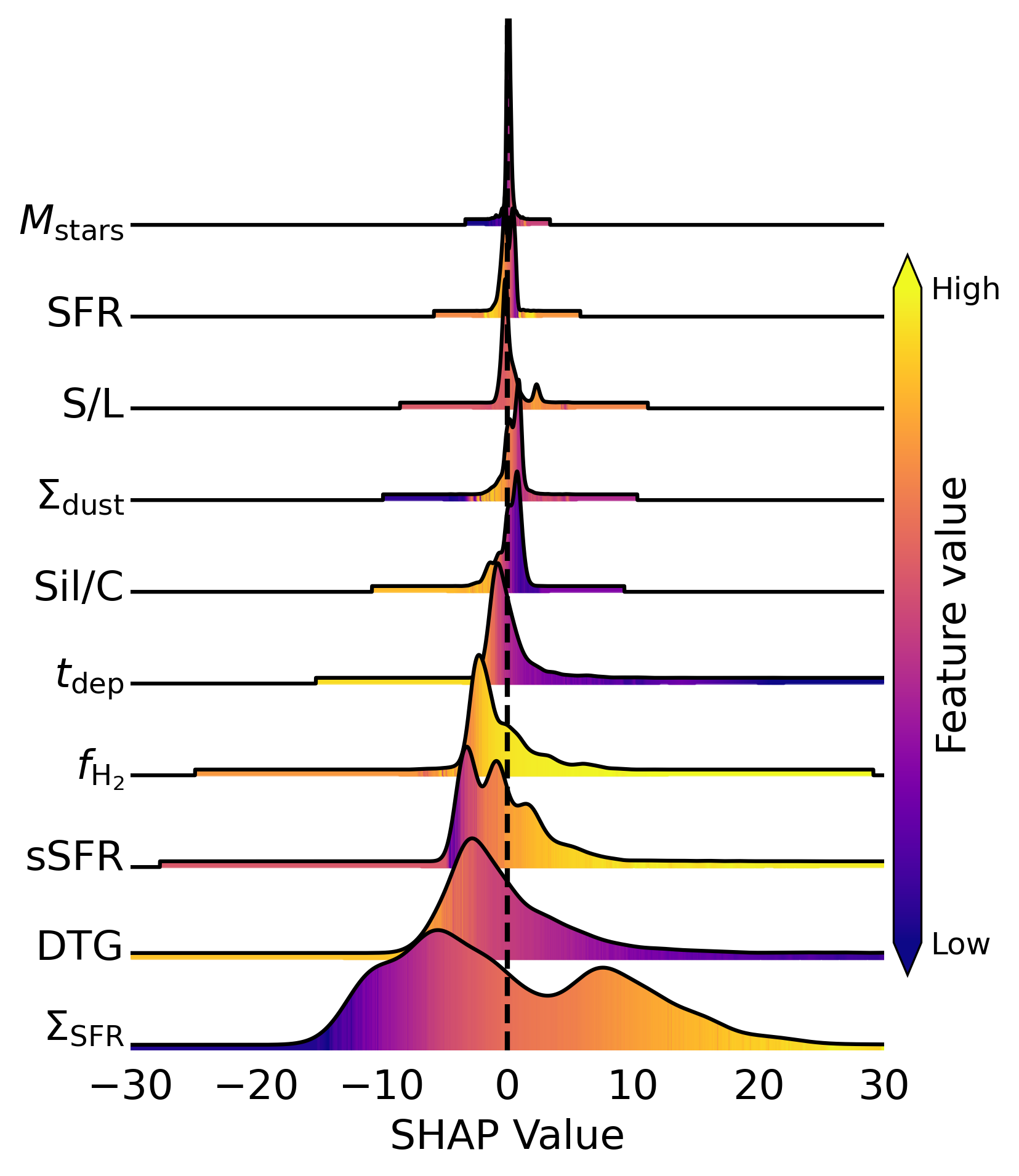}
    \caption{Ridgeline plot showing the distribution of SHAP values for each feature in predicting $T_{\rm dust}$ in our galaxy sample. The dashed vertical line marks zero. Histogram areas are colored according to the normalized values of each feature, from dark blue (low) to yellow (high).}
    \label{fig:shap}
\end{figure}

\begin{table}[]
\centering
\caption{Feature importance in predicting \tdust}
\label{tab:FI}
\begin{tabular}{cc}
\hline
Feature & Importance \\
\hline
$\Sigma_{\rm SFR}$ & $7.35$ \\
DTG & $3.62$ \\
sSFR & $2.42$ \\
$f_{\rm H_2}$ & $2.42$ \\
$t_{\rm dep}$ & $1.11$ \\
\hline
\end{tabular}
\caption*{\small Note: Feature importances are derived from the Shapley analysis as the mean of absolute values of SHAP values for each feature. Only features with importances larger than $1$ are shown.}
\end{table}


\begin{figure}[t]
    \centering

    \begin{subfigure}[b]{\columnwidth}
        \centering
        \includegraphics[width=\columnwidth]{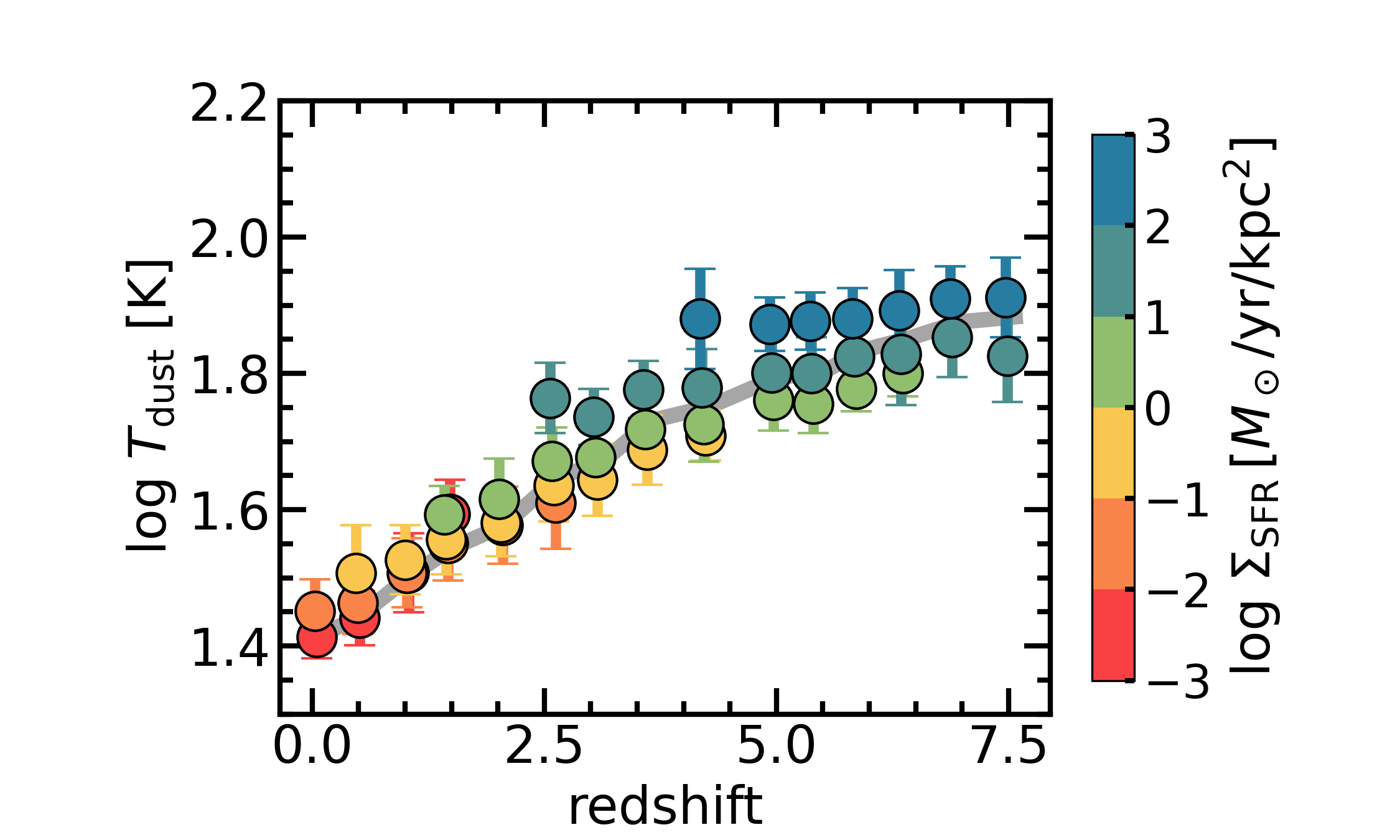}
    \end{subfigure}
    
    \vskip -0.4cm

    \begin{subfigure}[b]{\columnwidth}
        \centering
        \includegraphics[width=\columnwidth]{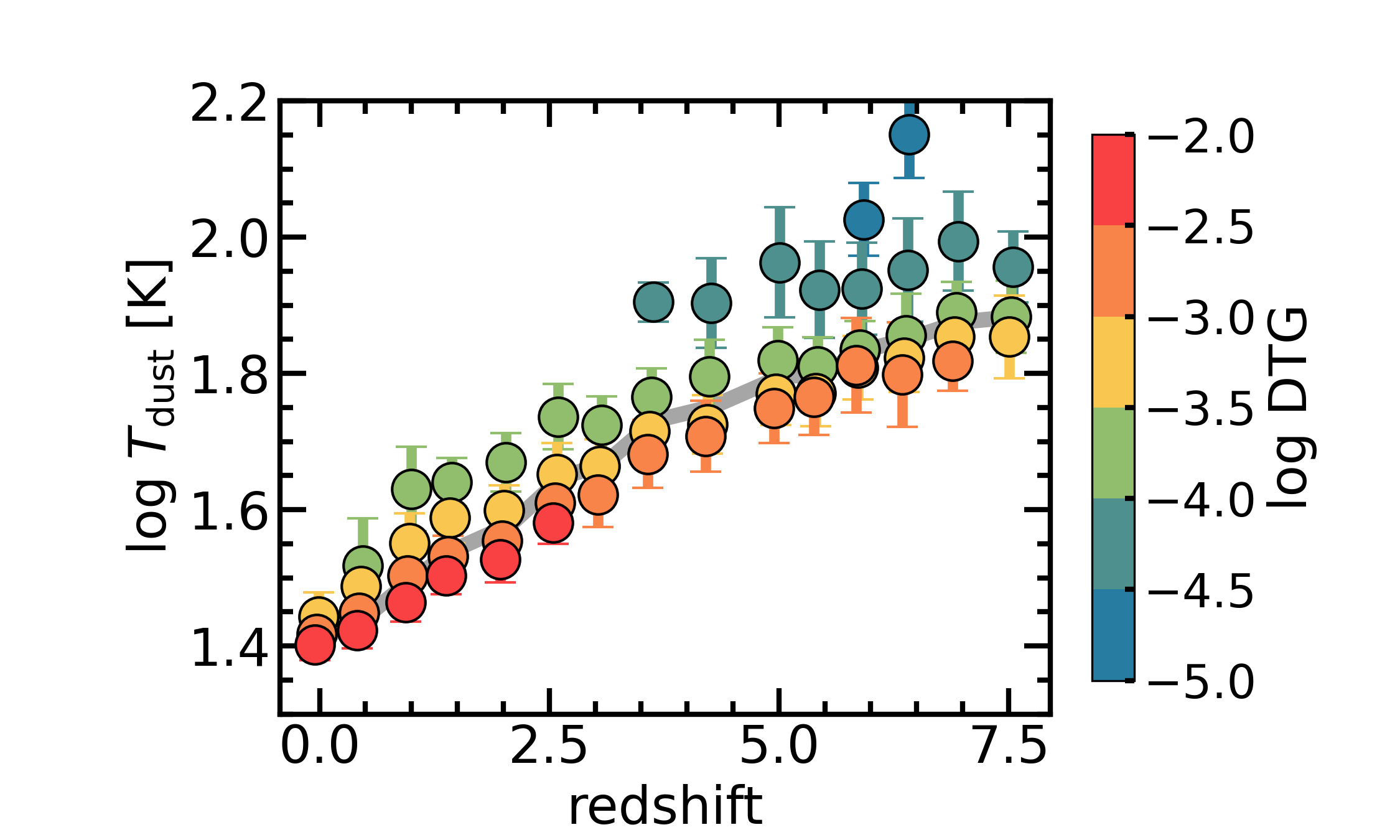}
    \end{subfigure}

    \caption{Dust temperature evolution with redshift separating galaxies in bins of SFR surface density ($\Sigma_{\rm SFR}$; top panel) and dust-to-gas ratio (DTG; bottom panel), as indicated by the colorbar. Each point and error bar indicates the median and $16$–$84$th percentiles of the distribution. Gray lines in the background refer to the full population.}
    \label{fig:Tdust_SFRDTG}
\end{figure}

We perform a Shapley analysis \citep{Shapley1953} to quantify the relative importance of different physical properties of galaxies in determining their dust temperature across redshift.
This technique -- already adopted in similar problems in astrophysics \citep[e.g.,][]{Gilda21, Narayanan26} -- originates from game theory and its original goal was to determine how team members cooperating toward a common goal should divide the resulting prize according to their individual contributions. Essentially, it assigns each feature a Shapley (or SHAP) value, representing the contribution of that feature to the model output. This method not only identifies which features most strongly influence the predictions, but also indicates the direction of their effects. Conceptually, a Shapley value measures how much a given feature shifts the model prediction away from its expected value when that feature is held fixed. 
Larger absolute SHAP values indicate stronger feature influence, while positive (negative) values correspond to an increase (decrease) in the predicted quantity. For example, SHAP values of $+20$ and $+10$ for features A and B, respectively, indicate that A has a stronger influence on the prediction, contributing twice as much as B. In our case, where the target is dust temperature, this corresponds to A inducing a $+20$ K increase in the predicted value relative to the prediction obtained when A is set to its mean (rather than its true) value.\\
In this work, we use the Shapley analysis to clarify the role of various galaxy physical properties in driving the dust temperature value, and in particular, its increase with redshift. The considered features (the \textit{players} in Shapley terminology) include the star formation rate (SFR), specific SFR ($\rm{sSFR}={\rm SFR}/M_{\rm stars}$), SFR surface density ($\Sigma_{\rm SFR}={\rm SFR}/A_{\rm gal}$)\footnote{The surface of the galaxy is estimated as $\pi R_{\rm gas}^2$, being $R_{\rm gas}$ the gas scale radius, consistently with what is used in the SED fitting procedure (Sect.~\ref{sec:dustsed}).}, gas depletion time ($t_{\rm dep} = M_{\rm gas}/{\rm SFR}$), stellar mass ($M_{\rm stars}$), dust mass ($M_{\rm dust}$), dust surface density ($\Sigma_{\rm dust}=M_{\rm dust}/A_{\rm gal}$), dust-to-gas mass ratio (DTG$=M_{\rm dust}/M_{\rm gas}$)\footnote{In the computation of the DTG, both molecular and atomic gas are considered, as both contribute to the cold gas reservoir of SAM galaxies.}, molecular gas fraction ($f_{H2}=M_{\rm H_2}/M_{\rm gas}$), and the mass ratios of small-to-large (S/L) and silicate-to-carbonaceous (Sil/C) grains.
These quantities are not only fundamental descriptors of galaxy evolution, but also trace the physical mechanisms that govern dust heating: the intensity of the interstellar radiation field (linked to SFR), the structural geometry of the galaxy (through surface densities), and the abundance and composition of dust.\\
The first step in performing the Shapley analysis is to construct a model capable of predicting the dust temperature for our galaxy sample. We trained a supervised regression model based on Extreme Gradient Boosting (XGBoost)\footnote{\url{https://xgboost.readthedocs.io/en/stable/}}, whose goal is to predict the target quantity (in our case $T_{\rm dust}$) from a set of input features. The sample of galaxies was randomly divided into training and test subsets using an 80/20 split. The model achieves an $R^2$ score\footnote{This is a metric indicating the quality of the model's performance, with $100\%$ indicating the best possible score (see \url{https://scikit-learn.org/stable/modules/generated/sklearn.metrics.r2_score.html}).} of $\approx 98\%$ on the training set and $\approx 93 \%$ on the test set, indicating good predictive accuracy and no significant overfitting.
We then applied the SHAP framework\footnote{\url{https://shap.readthedocs.io/en/latest/index.html}} to compute the SHAP values for each feature and interpret the contribution of individual features to the model outputs. The results are illustrated in Fig. \ref{fig:shap} as a ridgeline (or joy) plot\footnote{Ridge plots are colloquially known as \textit{joy plots} after the cover of Joy Division's Unknown Pleasures (1979). The cover image -- a visualization of pulses from CP 1919, the first observed pulsar (\citealt{pulsar}; Nobel Prize in Physics, 1974) -- was originally produced by \cite{craft1970} and later reproduced in The Cambridge Encyclopaedia of Astronomy (1977).}.
A quantitative summary of the most influential features is presented in Tab. \ref{tab:FI}, where the importance of each feature is expressed as the mean of the absolute SHAP values.\\
The Shapley analysis clearly identifies the SFR surface density as the main feature contributing to the value of the dust temperature: higher (lower) $\Sigma_{\rm SFR}$ values correspond to higher (lower) $T_{\rm dust}$, as shown by the colors in Fig. \ref{fig:shap}. Other relevant features in this analysis are the DTG ratio, sSFR, molecular gas fraction, and depletion time, with galaxies featuring lower DTG, higher star formation activity, and larger molecular gas fractions being associated with warmer dust, respectively. \\

The role of the two most important features, $\Sigma_{\rm SFR}$ and DTG, on $T_{\rm dust}$ is further illustrated in Fig. \ref{fig:Tdust_SFRDTG}, which shows the redshift evolution of dust temperature across different bins of these quantities. A similar analysis for the three other properties listed in Tab. \ref{tab:FI} is presented in App. \ref{app:tdust_rel}. The adopted binning scheme highlights that galaxies at higher redshift reach significantly larger star formation surface densities, up to $\gtrsim 10^{2}\,M_\odot\,{\rm yr^{-1}\,kpc^{-2}}$, and lower DTG ratios, down to $\lesssim 3\times10^{-5}$, both of which contribute to higher dust temperatures.\\
The relevance of $\Sigma_{\rm SFR}$ stands in its ability to capture two key factors driving dust heating: the intensity of the UV radiation field (traced by the SFR) and the spatial geometry (traced by the size of the galactic disk). More compact star-forming galaxies feature higher radiation energy densities, leading to more efficient heating of dust grains. This is mostly relevant to diffuse dust, which is distributed throughout the galactic disk.
\\
The relevance of the DTG is twofold. First, galaxies with a lower DTG contain fewer dust grains, which results in more efficient heating for a given SFR. Second, the DTG is relevant for the radiative transfer in dense molecular clouds. These systems are optically thick, often featuring self-absorption of IR photons emitted by dust. The optical depth is determined by DTG (and by the $M_{\rm MC}/R^2_{\rm MC}$ ratio, a parameter of the RT model), with lower DTGs corresponding to lower optical depths. In these cases, inner shells of the MCs become more transparent and can thus contribute to IR emission, which is hotter than the diffuse one. \\
In conclusion, the increase of $T_{\rm dust}$ with redshift is mainly driven by the prevalence, in the early Universe, of galaxies with higher SFR surface densities. A secondary, less dominant effect -- according to our Shapley analysis -- comes from the enhanced contribution of lower DTGs. We also note that dust properties -- namely, size and chemical composition of grains -- do not play a relevant role.

\subsection{Temperature scaling relations}
\label{sec:scalingscatter}

\begin{figure}[t]
    \centering

    \begin{subfigure}[b]{\columnwidth}
        \centering
        \includegraphics[width=\columnwidth]{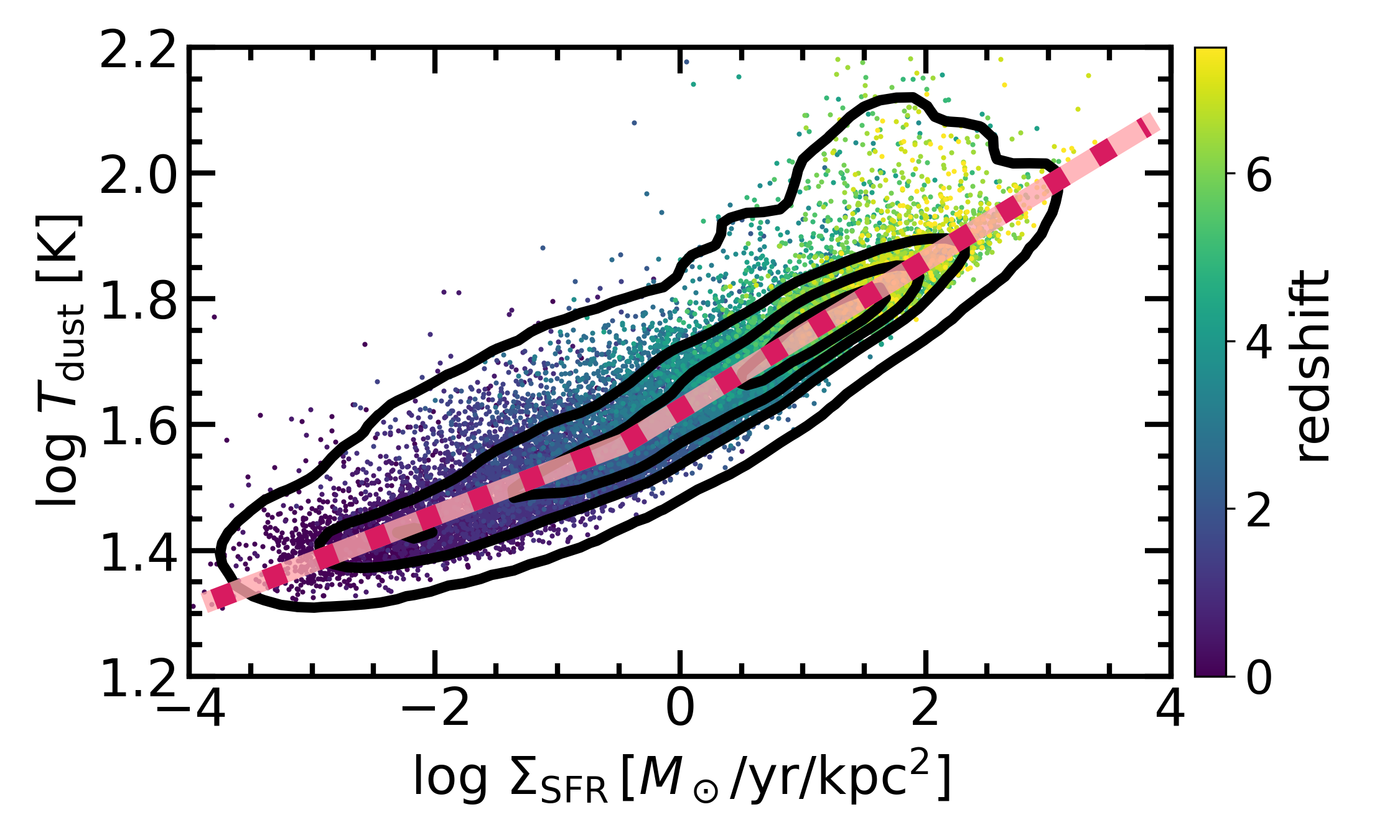}
        \label{fig:Tdust_SigmaSFR}
    \end{subfigure}

    \begin{subfigure}[b]{\columnwidth}
        \centering
        \includegraphics[width=\columnwidth]{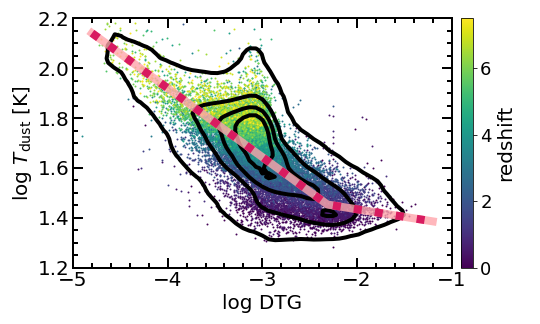}
        \label{fig:Tdust_DTG}
    \end{subfigure}

    \caption{Dust temperature as a function of $\Sigma_{\rm SFR}$ and DTG. Each point represents a galaxy and is color-coded by redshift, while the black contours correspond to the full sample. The red dashed line shows a broken power law fit to data (Tab. \ref{tab:coefffit}).}
    \label{fig:Tdust_scaling}
\end{figure}

\begin{table}[]
\centering
\caption{Best-fit parameters describing the relation between \tdust and various galaxy properties.}

\begin{tabular}{lccccc}
\hline
$x$ & $a$ & $b$ & $x_{\rm crit}$ & $q$ & $\sigma_{\rm res} [{\small \rm dex}]$\\
\hline
$\Sigma_{\rm SFR} $ & $0.070 $ & $0.11 $ & $-0.63 $ & $1.6$ & $0.06$\\
DTG  & $-0.28$ & $-0.060 $ & $-2.3$ & $0.81$ & $0.10$ \\
sSFR  & $0.096$ & $0.30 $ & $-9.3$ & $2.4$ & $0.06$\\
$f_{\rm H_2}$  & $0.22$ & $0.96 $ & $-0.39$ & $1.6$ & $0.09$ \\
$t_{\rm dep}$  & $-0.26$ & $-0.033 $ & $0.80$ & $1.6$ & $0.10$ \\

\hline
\end{tabular}
\label{tab:coefffit}
\caption*{\small Notes: The fit relations are broken power laws (Eq. \ref{eq:broken}) where $y = {\rm log \,}T_{\rm dust} [{\rm K}]$ and $x = \log \Sigma_{\rm SFR} [M_\odot / {\rm yr} / {\rm kpc^2}]$, $\log {\rm DTG}$, $\log {\rm sSFR \, [yr^{-1}]}$, $\log f_{\rm H_2}$, or $\log t_{\rm dep} \, [{\rm Gyr}]$. The $\sigma_{\rm res}$ column gives the standard deviation of the residuals for each relation, which is a proxy of the scatter of the data points around the best-fit relation.}
\end{table}

Figure \ref{fig:Tdust_scaling} shows the dependence of galaxy dust temperature on the two properties found to have the strongest impact, namely the SFR surface density and DTG ratio. Analogous relations for the three other influential properties listed in Tab. \ref{tab:FI} are presented in App. \ref{app:tdust_rel}. Dust temperature increases systematically with increasing $\Sigma_{\rm SFR}$ and with decreasing DTG. To provide a convenient, ready-to-use parametrization of these trends, we fit the relations with a broken power law of the type

\begin{equation}
y(x) =
\begin{cases}
a \, x + q, & x < x_{\rm crit} \\[2mm]
b \, (x - x_{\rm crit}) + \big(a \, x_{\rm crit} + q\big), & x \ge x_{\rm crit}.\\
\end{cases}
\label{eq:broken}
\end{equation}
We adopt the broken power-law model rather than a simple linear fit to account for the flattening of the relation at low redshifts. The best-fit coefficients -- both for this relation and for those shown in Fig. \ref{fig:Tdust_scaling2} -- are listed in Table \ref{tab:coefffit}.\\
As indicated by the critical values of the broken power-law fits, both relations show a flattening at $z \lesssim 2$, occurring for relatively high DTG ratios ($\gtrsim 5 \cdot 10^{-3}$) and low $\Sigma_{\rm SFR} (\lesssim 0.1\,M_\odot\,{\rm yr^{-1}\,kpc^{-2}}$). The steeper slopes observed at higher redshifts could partially explain the evolution of the \tdust scatter discussed in Sect.~\ref{sec:tdust_evol_z}: at high $z$, even small variations in DTG or $\Sigma_{\rm SFR}$ produce significant changes in $T_{\rm dust}$, whereas at low $z$ the relation flattens.\\
Given the clear dependence of the dust temperature on $\Sigma_{\rm SFR}$ and DTG, we use our results to provide a relation that allows estimation of the DTG from $\Sigma_{\rm SFR}$, $T_{\rm dust}$, and redshift. This is particularly useful for observational studies, where $\Sigma_{\rm SFR}$ and $T_{\rm dust}$ are typically derived from SED fitting, while DTG is more challenging to determine. Direct DTG measurements usually require gas mass estimates from line detections \citep[e.g.,][]{Vallini18}, which are often highly uncertain \citep[e.g.,][]{Zanella18, Salvestrini25}, or rely on knowledge of the ISM metallicity to exploit the DTG-$Z_{\rm gas}$ relation \citep[e.g.,][]{RR14, Popping22, Algera26}.\\
We adopt a relation of the form:

\begin{equation}
{\rm log \, DTG} = A \, {\rm log} \left(\frac{T_{\rm dust}}{\rm K}\right) + B \, {\rm log} \left(\frac{\Sigma_{\rm SFR}}{{\rm M_\odot \, yr^{-1} \, kpc^{-2}}}\right) + C \, {\rm log}(1+z) + D\,
\end{equation}

with $A=-4.129 \pm 0.027$, $B=0.456 \pm 0.003$, $C=-1.295 \pm 0.019$, and $D=4.521 \pm 0.042$. Comparing the DTG values predicted by this relation with the true values shows good, though not perfect, agreement ($R^2 \approx 0.78$), with a standard deviation of residuals of $\approx 0.2 \, {\rm dex}$. 
Although including additional variables or adopting more sophisticated -- potentially machine-learning–based -- parametrizations could improve performance, we opt for a simpler approach. This choice is motivated by the fact that large uncertainties in standard high-$z$ gas conversion factors \citep[e.g.,][]{FriasCastillo25, Tripodi24b, Decarli25} introduce errors comparable to, or greater than, those arising from the simple parametrization.

\subsection{Dust emissivity index}
\label{sec:betadust}

\begin{figure}
    \centering
    \includegraphics[width=0.99\columnwidth]{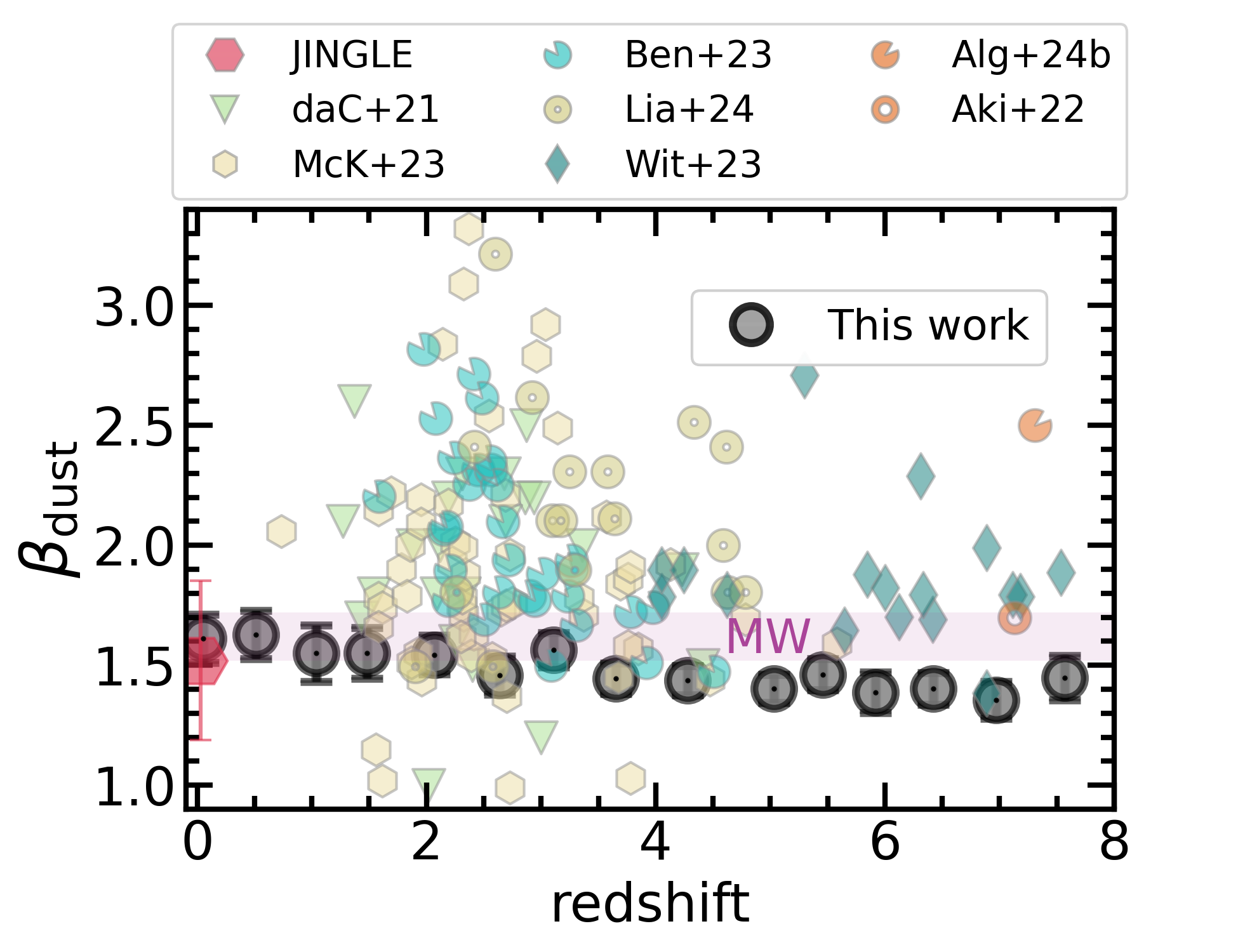}
    \caption{Dust emissivity index evolution $\beta_{\rm dust}$ with redshift. Points and error bars refer to median and $16-84$th percentiles range. Some observational measurements are shown for reference: the local JINGLE sample \citep{Lamperti19} and individual higher-$z$ sources from \citet{daCunha21, McKay23, Bendo23, Liao24, Witstok23, algera2024b, Akins22}. The effective Milky Way value of $\beta_{\rm dust}=1.62\pm0.10$ is shown for reference as a purple area \citep{Galliano18}.
    We remind that the intrinsic emissivity index of grains as assumed by our model is $\beta_{\rm dust}=2$. Therefore, the observed flattening at high-redshift is likely due to multi-temperature dust, and not to variations in dust properties.}
    \label{fig:beta_z}
\end{figure}

We present in Fig. \ref{fig:beta_z} the redshift evolution of the dust emissivity index $\beta_{\rm dust}$ as predicted by our model, which slightly increases toward low redshift. 
Owing to the strong anti-correlation between $\beta_{\rm dust}$ and \tdust in the SED fitting procedure, part of the observed redshift trend reflects the increase of \tdust with redshift discussed in Sect. \ref{sec:tdust_evol_z}. The dominant role of \tdust in driving variations of $\beta_{\rm dust}$ is further confirmed by a Shapley analysis analogous to that presented in Sect.~\ref{sec:shap}, but with the emissivity index as the target variable (not shown here for brevity). In this case, \tdust is the most influential feature, with an importance more than a factor of $\gtrsim 3$ higher than that of any other parameter. At the same time, removing \tdust from the set of input features in the regression model reduces the test-sample coefficient of determination from $87\, \%$ to $78\,\%$, highlighting once again the impact of the SED fitting–induced anti-correlation between these two quantities.\\
With this caveat in mind, we turn to a more physical speculation. Although \betadust is often linked to the chemical composition of dust grains, \textsc{GRASIL} adopts the classical grain absorption cross sections computed by \citet{Laor93}, based on exact Mie theory for spherical grains. As a consequence, at wavelengths much larger than the grain size (typically longwards a few tens of microns and longer), the intrinsic emissivity index of individual grains is $\beta_{\rm dust}=2$.\\
Therefore, any deviation from this value does not arise from changes in the microscopic dust properties, but rather from the superposition of dust grains at different temperatures (e.g., \citealt{Hunt15}, \citealt{Sommovigo&Algera25}). These temperatures depend on both the local radiation field and the grain size and composition. This temperature mixing flattens the far-IR decline of the SED, mimicking the effect of a lower emissivity index when the SED is fitted with a single-temperature MBB.
The analysis in Sect.~\ref{sec:shap} suggests such a temperature mixing to be more pronounced at high redshift, where the contribution of warm molecular clouds is enhanced due to lower DTG ratios, resulting in a broader overall temperature distribution.\\
This point should also be kept in mind when interpreting values of \betadust (and $T_{\rm dust}$) inferred under the simplifying assumption of a single dust temperature for the entire galaxy. Moreover, laboratory measurements suggest that in real grains \betadust depends on, and in particular anti-correlates with, their temperature \citep[e.g.,][]{Coupeaud11}.\\
Our model predicts a mild redshift evolution toward $\beta_{\rm dust} \approx 1.6$ at low redshift. When comparing with observational estimates inferred from SED fitting, the predicted trend is broadly consistent with the local JINGLE sample \citep{Lamperti19}, but observed values at all redshifts span a much wider range, $\beta_{\rm dust} \approx 1-3$, with no clear redshift dependence -- a scatter substantially larger than our simulations produce. The discrepancy is most pronounced at high redshift \citep{daCunha21, Witstok23, McKay23, Bendo23, Liao24, algera2024b, Akins22}.\\
Several observational systematics may contribute to this scatter, including prior-dependent degeneracies in SED fitting \citep[e.g.,][]{Witstok23} -- exacerbated by the sparse far-IR photometric coverage typical of high-redshift sources  -- and assumptions regarding optical depth \citep[e.g.,][]{McKay23}. On the model side, temperature mixing lowers the effective emissivity index. If this effect is more pronounced in simulations than in the observed galaxies, it could account for part of the discrepancy. At the same time, the highest inferred values ($\beta_{\rm dust} \gtrsim 2.5$) are difficult to reconcile with theoretical dust models, which typically predict $\beta_{\rm dust} \approx 1-2$ \citep[e.g.,][]{DraineLee84, Kohler15}. Grain coagulation or other modifications to optical properties may be required to explain such steep values \citep[see][and discussion in \citealt{algera2024b}]{Kohler15}. It is also worth noting that limited sampling of the observed far-infrared SED, particularly in high-redshift studies where observations are often time-consuming and sparse, can significantly affect the determination of the dust emissivity index.\\
Finally, we verified that adopting a fixed value of \betadust$=1.6$, as measured in the Milky Way \citep{Galliano18} and consistent with the range commonly assumed in observational studies (e.g., see App. \ref{app:obs_data}), does not significantly impact the inferred evolution of dust temperature with redshift. The resulting $T_{\rm dust}$ values are fully consistent with the median trend shown in Fig. \ref{fig:Tdust_z}, showing an almost negligible difference up to $z<2$ and only a modest reduction in the median \tdust\ at higher redshift ($\Delta T_{\rm dust}\approx4$ K).

\section{Summary}
\label{sec:conclusions}

In this paper, we have investigated the redshift evolution of dust temperature as predicted by a semi-analytic cosmological model of galaxy evolution that includes an explicit treatment of dust production and evolution \citep{Parente2023}, coupled with a radiative transfer pipeline to model the IR emission from dust grains (\textsc{GRASIL}; \citealt{Silva98}). To reproduce as closely as possible the observational determination of dust temperature, we derive \tdust by fitting the resulting SEDs with a single-temperature MBB with the \textsc{EOS-Dustfit} code \citep{Tripodi24b}, under the optimistic assumption of full spectral coverage.\\
Our main results can be summarized as follows.
\begin{itemize}
\item We find a clear increase in $T_{\rm dust}$ with redshift, in good agreement with the overall trend suggested by observational constraints, despite the large scatter present in the data.

\item Using a Shapley analysis, we identify the star formation rate surface density $\Sigma_{\rm SFR}$ and the dust-to-gas (DTG) ratio as the primary drivers of the increase in \tdust toward high redshift, where galaxies host more compact and intense star formation and are more dust poor.

\item The role of $\Sigma_{\rm SFR}$ can be understood in terms of higher star formation surface densities being associated with stronger radiation fields, leading to enhanced dust heating. Lower DTGs correspond to more optically thin molecular clouds -- which appear hotter -- and to a smaller number of dust grains over which stellar energy is distributed, resulting in increased heating per grain.

\item The relations between \tdust and $\log \Sigma_{\rm SFR}$ or $\log {\rm DTG}$ show a progressively steeper dependence with increasing redshift, which we parametrize with a broken power law. We also present a simple relation to derive DTG from $\Sigma_{\rm SFR}$, $T_{\rm dust}$, and redshift, which is particularly useful in observational studies where direct gas mass measurements are not feasible.

\item We predict a mild redshift evolution of $\beta_{\rm dust}$, driven by temperature mixing. Observed values span a much wider range with no clear redshift trend, with the highest inferred values ($\beta_{\rm dust}\gtrsim2.5$) difficult to reconcile with current grain models.\\
\end{itemize}
By combining, for the first time, a cosmological galaxy evolution model with explicit treatment of dust physics, detailed radiative transfer, and an observationally motivated strategy to infer $T_{\rm dust}$, we have shown that the main drivers of warm dust temperatures in the early Universe are the star formation rate surface density and the DTG ratio. Early galaxies are generally more compact and optically thinner, which naturally leads to warmer dust. This finding, consistent with previous theoretical studies \citep[e.g.,][]{Ma2019, Liang19, Hirashita22, Vijayan22}, is especially important for interpreting dust properties in high-redshift galaxies, where far-IR emission is often incompletely sampled.\\
From an observational point of view, our findings are in good agreement with some results in the literature. \citet{Yan16} showed that the $L_{\rm IR}-T_{\rm dust}$ relation in dusty star-forming galaxies is a direct consequence of the compact size ($R_{\rm eff} \lesssim 2$ kpc) of their star-forming regions, providing direct evidence for the role of geometry in setting dust temperature. \citet{Shivaei22} further showed, through a two-temperature MBB fitting of $z\sim2.3$ galaxies, that the IR SED becomes warmer and broader with increasing $\Sigma_{\rm SFR}$, and that sub-solar metallicity -- and, hence, low DTG \citep[e.g.,][]{Popping22} -- galaxies display significantly hotter and broader IR SEDs than their solar-metallicity counterparts, pointing to an independent role of the DTG in shaping $T_{\rm dust}$. On the other hand, it has been suggested that the observed redshift evolution of $T_{\rm dust}$ may be partly an observational artifact, driven by the increase of SFR and IR luminosity at fixed stellar mass combined with a selection bias toward high-luminosity systems \citep[e.g.,][]{Dudze20,Drew22}. While such a bias can certainly play a role, our model indicates that the evolution is not purely a selection effect: physical changes in SFR, galaxy size, and ISM opacity each contribute independently and are sufficient to drive the trend even in the absence of observational bias.\\ 
It is nevertheless important to emphasize that the RT treatment adopted here -- based on a two-phase ISM model consisting of diffuse gas and molecular clouds -- is necessarily simplified due to the limited spatial resolution of the SAM. While this framework is made physically plausible (for example, by including a finite escape time for young stellar populations from dense molecular clouds; \citealt{Silva98}), it remains an oversimplification of the complex and irregular geometries of real galaxies. In our simulations, the gas component is imposed by construction to follow a disk-like configuration, a limitation that becomes particularly relevant at high redshift, where well-defined disk structures may not yet be in place \citep[but see also][]{Ferreira23, Kartaltepe23, Sun24}. Given the critical role of geometry in RT calculations, this constitutes a significant caveat of the model. A more realistic description would require full 3D hydrodynamic simulations, although producing large cosmological samples comparable to those generated by semi-analytic models remains challenging.\\
A further point worth discussing concerns the observationally motivated derivation of $T_{\rm dust}$. As already mentioned in Sect. \ref{sec:sedbin}, we have optimistically assumed full spectral coverage in order to marginalize over possible instrumentation-driven limitations and keep the focus on the physics of dust. However, current facilities do not allow observations of the rest-frame mid-IR emission of galaxies at $z \gtrsim 2$, where hot dust associated with star-forming regions or AGN may dominate the SED, potentially revealing a different -- and often suggested bimodal \citep[e.g.,][]{Liang19} -- view of dust temperatures in galaxies. This underscores the importance of possible future facilities such as the \textit{PRobe far-Infrared Mission for Astrophysics} ({PRIMA}; e.g., \citealt{Bradford22}) in shedding new light on this topic.\\

\begin{acknowledgements}
We thank the referee Hiddo Algera for the prompt report and constructive comments which improved the clarity and the quality of the work. MP was funded by NASA Astrophysics Theory Program grant 80NSSC24K1223 (PI: DN).  FS and CF acknowledge financial support from the PRIN MUR 2022 2022TKPB2P - BIG-z, Ricerca Fondamentale INAF 2023 Data Analysis grant 1.05.23.03.04 ``ARCHIE ARchive Cosmic HI \& ISM  Evolution'' , Ricerca Fondamentale INAF 2024 under project 1.05.24.07.01 MINI-GRANTS RSN1 "ECHOS", Bando Finanziamento ASI CI-UCO-DSR-2022-43 CUP:C93C25004260005 project ``IBISCO: feedback and obscuration in local AGN''.  DN is grateful for support from the NASA Astrophysics Theory Program via grants 80NSSC24K1223 and 80NSSC22K0716. GLG acknowledges financial support from the European Union's HORIZON-MSCA-2021-SE-01 Research and Innovation Programme under the Marie Sklodowska-Curie grant agreement number 101086388 - Project LACEGAL.

\end{acknowledgements}


\bibliographystyle{aa} 
\bibliography{aa59707-26.bib} 


\appendix

\section{Optically thick assumption in SED fitting}
\label{app:opt_thick}

\begin{figure}[t]
    \centering

        \begin{subfigure}[b]{\columnwidth}
        \centering
        \includegraphics[width=0.85\columnwidth]{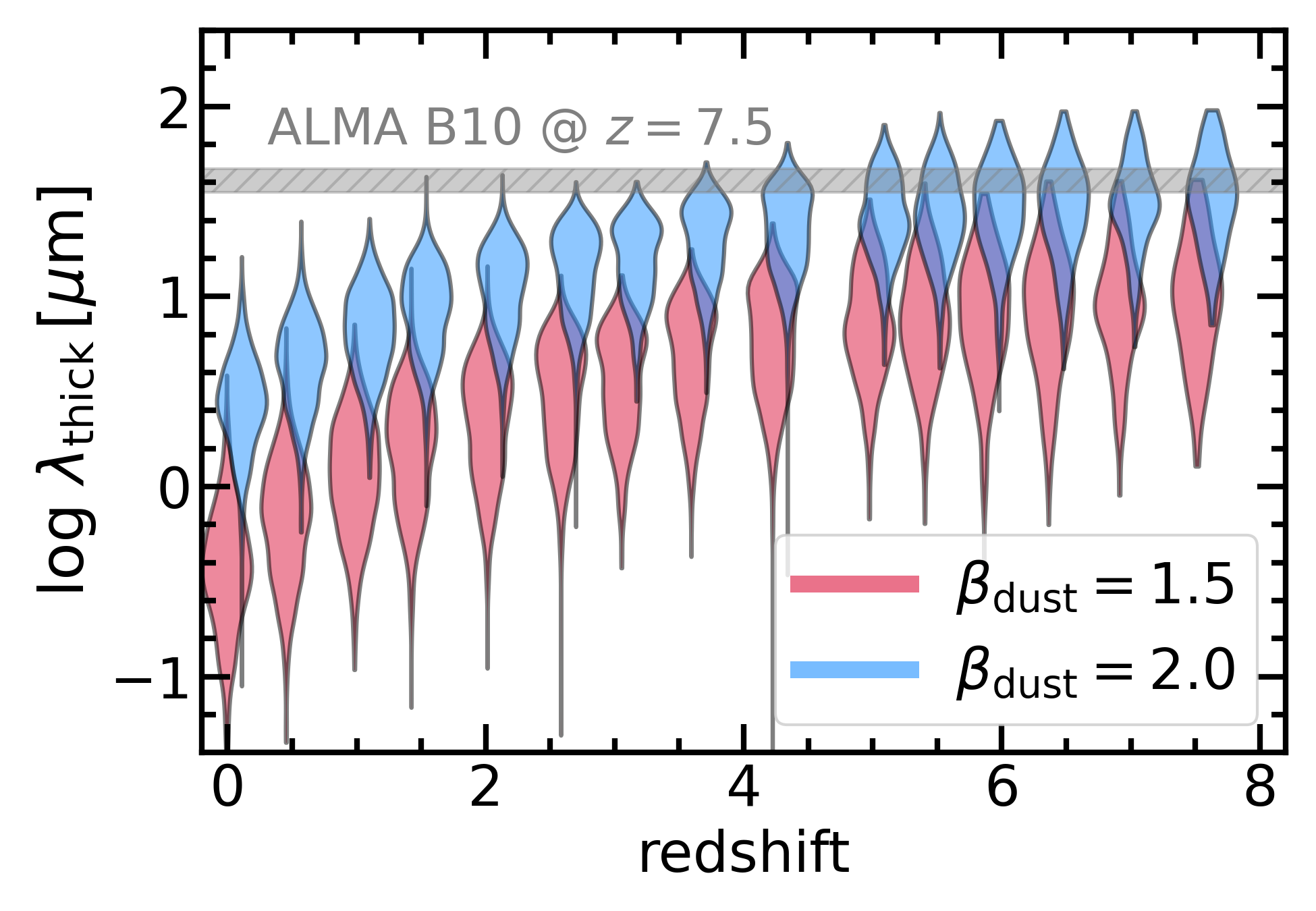}
    \end{subfigure}


      \begin{subfigure}[b]{\columnwidth}
        \centering
        \includegraphics[width=0.85\columnwidth]{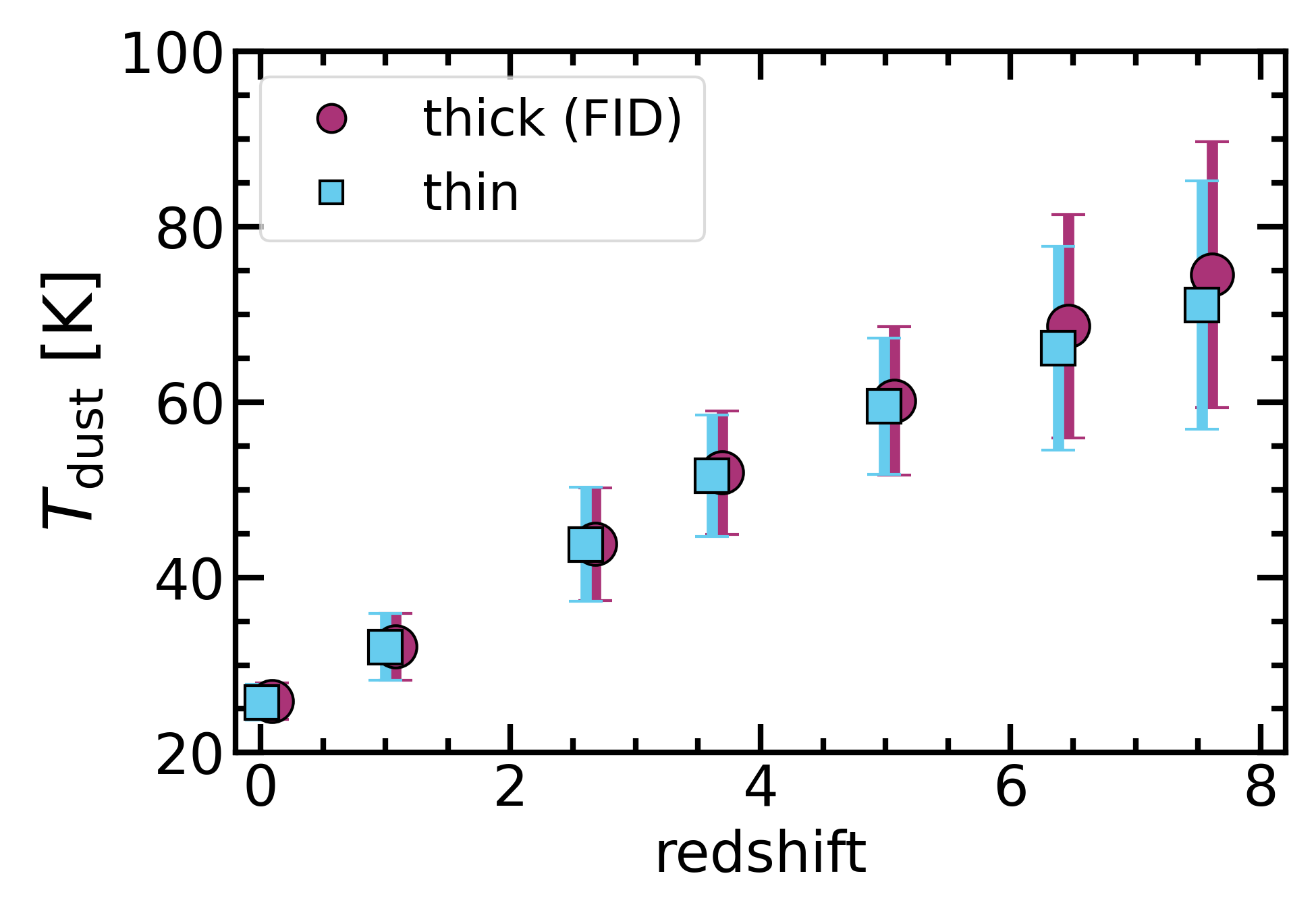}

    \end{subfigure}  

    \caption{Top panel: redshift evolution of $\lambda_{\rm thick}$, defined as the wavelength at which the optical depth becomes unity (Eq. \ref{eq:tau}). The violin plots show the distribution of $\lambda_{\rm thick}$ for our galaxy sample assuming $\beta_{\rm dust}=1.5$ and $2$. For reference, we also indicate the rest-frame wavelength corresponding to ALMA band 10 at $z=7.5$, the highest redshift considered in this work. \\
    Bottom panel: evolution of dust temperature with redshift assuming either an optically thick medium (our fiducial model; purple circles) or an optically thin medium (cyan squares). Symbols indicate the median values, while error bars show the $16$th--$84$th percentile range of the distributions.}
    \label{fig:thin}
\end{figure}

In our SED fitting procedure, we assume an optically thick medium, introducing the factor $1-e^{-\tau_\nu}$ in Eq. \ref{eq:Snufit}. Here we show that this assumption has only a minor impact on our results. This test is motivated by the fact that observational studies often adopt the optically thin approximation, which can bias the inferred dust temperatures toward lower values (e.g., \citealt{daCunha21, algera2024}).\\
The validity of the optically thin approximation, i.e. $\tau_\nu \ll 1$, mainly depends on the compactness of the dust distribution, as it is clear from Eq. \ref{eq:tau}. By inverting this relation and fixing $\beta_{\rm dust}$, we can estimate the transition wavelength between the optically thin and thick regimes, defined by $\tau_\nu \approx 1$. The top panel of Fig. \ref{fig:thin} shows the resulting $\lambda_{\rm thick}$ for both $\beta_{\rm dust}=1.5$ and $2$.
Because galaxies (hence the dust distribution) become increasingly compact toward higher redshift, the medium remains optically thick up to progressively larger wavelengths. However, comparing the rest-frame $\lambda_{\rm thick}$ with the (redshifted) highest-frequency ALMA band accessible at $z\approx7.5$ (the highest redshift considered in our catalogs), we find only a marginal overlap. This indicates that the current observational setup adopted in this work is only marginally able to probe the wavelengths where the medium is -- in theory -- optically thick. This suggests that the optically thin approximation is largely valid.

To verify this explicitly, we repeat the SED fitting analysis assuming an optically thin medium, replacing the $1-e^{-\tau_\nu}$ term in Eq. \ref{eq:Snufit} with $\tau_\nu$. 
The resulting dust temperature evolution with redshift is shown in the bottom panel of Fig. \ref{fig:thin}. The differences with respect to the optically thick case are generally minor, becoming only marginally significant at the highest redshifts. In this regime, the optically thin assumption can lead to inferred dust temperatures lower by up to a factor of $\approx 1.2$. We therefore conclude that, given the current observational setup, the optically thin approximation remains a reasonable assumption according to our model predictions.

\section{Observational data}
\label{app:obs_data}
In this section, we briefly summarize the properties of the observational samples included in Fig.~\ref{fig:Tdust_z} and discussed in Sect.~\ref{sec:tdust_evol_z}.
We compiled $T_{\rm dust}$ measurements from several literature studies spanning a wide range in redshift ($0\lesssim z\lesssim8.5$) and galaxy populations. 
For details about the selection criteria adopted and the analysis performed in each of the literature works, we refer to the references listed in this section.\\ 
We first consider works that report $T_{\rm dust}$ measurements for star-forming galaxies, both in the local Universe \citep[e.g.,][]{Clements18, Lamperti19, Nersesian19} and at higher redshift \citep[e.g.,][]{Schreiber18, Franco20, Viero22, Witstok23}. Most of the samples at intermediate and high redshift ($z>1$) are dominated by dusty star-forming galaxies and submillimeter galaxies (DSFGs and SMGs, respectively; e.g., \citealt{Faisst20, Bakx21, Witstok23, algera2024, algera2024b, Bing26}), whose optical and UV emission is largely obscured by dust and whose bolometric output is predominantly emitted at far-IR and submillimeter wavelengths. The sample presented by \citet{Bakx24} instead consists of [C\,\textsc{ii}]-emitting galaxies detected as companions of luminous quasars.
In the highest redshift regime ($z\gtrsim7$), the galaxies included in our compilation are predominantly classified as Lyman-break galaxies \citep[e.g.,][]{Bakx20a, Bakx25, Mitsuhashi24, Sugahara21, Witstok22b}. Some of the considered samples also include galaxies whose observed brightness is significantly enhanced by gravitational lensing \citep[e.g.,][]{Bakx20b, Reuter20}.\\
We also included a few samples of ultra-luminous infrared galaxies (ULIRGs), mainly in the redshift range $0\lesssim z\lesssim2.5$ \citep[e.g.,][]{Yang07, Younger09, Magdis10, Magdis14, Paspaliaris21}. These systems are characterized by larger SFRs and $\Sigma_{\mathrm{SFR}}$ \citep[e.g.,][]{Hogan22}, compared to galaxies lying on the star-forming main sequence \citep[e.g.,][]{Renzini15}, and are therefore expected to exhibit higher $T_{\rm dust}$ as a consequence of their intense and compact star formation activity \citep[e.g.,][]{Witstok23}.\\
Although our simulation does not explicitly include Active Galactic Nuclei (AGNs), we incorporated several observational samples of high-redshift AGN and QSOs \citep[e.g.,][]{Duras17, Bischetti18, Bischetti21, Tripodi24b, Costa26, FernandezAranda25}. AGN, and quasars in particular, can contribute significantly to dust heating even on kiloparsec scales (e.g., \citealt{Tsukui23, FernandezAranda25}), potentially shifting the far-IR SED toward shorter wavelengths and yielding higher $T_{\rm dust}$ values when inferred using a single MBB. However, this effect does not appear to be systematic in the sources included in our compilation, suggesting that the combined contribution of AGN and stellar radiation to dust heating does not necessarily result in elevated $T_{\rm dust}$ values.
We also note that the identification of AGN activity can be far from straightforward, especially in systems with relatively low AGN luminosities ($L_{\rm bol}\approx10^{43\text{--}44}\,{\rm erg\,s^{-1}}$) or in heavily obscured objects. As a result, some of the ULIRGs, DSFGs, and SMGs included in our compilation may host a dust-obscured AGN in their central regions \citep[e.g.,][]{Uematsu25, Kiyota26}.\\
Since we used a fit with a single-temperature MBB to derive $T_{\rm dust}$, we preferred the works in the literature that fitted the far-IR SED with a single-temperature MBB. In the case of multiple methods adopted in a single study, we considered the $T_{\rm dust}$ measurements derived with a methodology similar to that adopted in Sect. \ref{sec:dustsed}. 
Some of the studies we included in Fig. \ref{fig:Tdust_z} adopted a fixed $\beta_{\rm dust}$ as they do not have enough photometric points in the Rayleigh–Jeans regime, to limit the degeneracy between $\beta_{\rm dust}$ and $T_{\rm dust}$. In particular, the more common assumptions were $\beta_{\rm dust} = 1.5$ \citep{Younger09, Magdis10, Magdis14, Franco20, Sugahara21, Witstok22b, Bakx24, Mitsuhashi24}, $\beta_{\rm dust} = 1.6$ \citep{Duras17}, and $\beta_{\rm dust} = 2$ \citep{Bakx20b, Bakx24b, Reuter20}.
\cite{Franco20} presented the results of the fit of the far-IR SED with a MBB on a sample of 35 galaxies at intermediate redshift, and discussed that when comparing the results obtained with $\beta_{\rm dust} = 1.5$ and $\beta_{\rm dust} = 2$, in the latter case $T_{\rm dust}$ values are $\approx4$~K lower, on average.\\
We remind the reader that \cite{Sommovigo21} introduced an alternative methodology to estimate $T_{\rm dust}$ by combining a single dust-continuum detection with the luminosity of  [C\,\textsc{ii}] at $158 \, \mu {\rm m}$ emission line -- a proxy of the gas mass. The same methodology was then applied also to samples of high-redshift dusty star-forming galaxies as those in the ALMA Large Program to INvestigate [C\,\textsc{ii}] at Early times (ALPINE; \citealt{Ferrara22}) and to the REBELS (Reionization Era Bright Emission Line Survey; \citealt{Schaerer20}). The results were presented in \cite{Sommovigo22b} and \cite{Sommovigo22a}, respectively.
We checked the consistency of the [C\,\textsc{ii}]-based approach with the single MBB model, and we found broadly consistent results for the sources at high-z from ALPINE \citep{Faisst20} and REBELS \citep{algera2024b, algera2024, Algera25c} presented in Fig.~\ref{fig:Tdust_z}.

\section{\tdust dependence on additional galaxy properties}
\label{app:tdust_rel}

\begin{figure}[t]
    \centering

    \begin{subfigure}[b]{\columnwidth}
        \centering
        \includegraphics[width=\columnwidth]{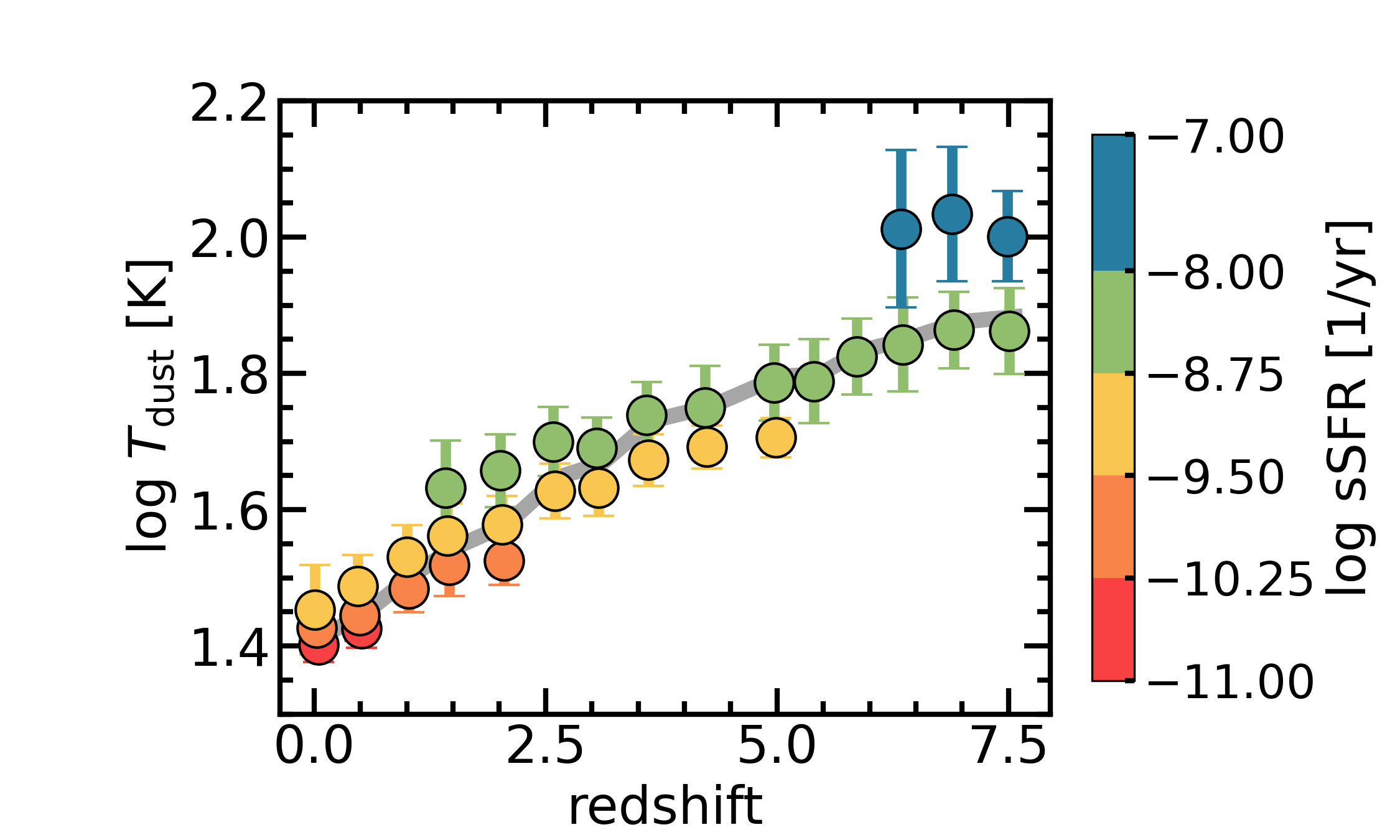}
    \end{subfigure}
    
    \vskip -0.4cm

    \begin{subfigure}[b]{\columnwidth}
        \centering
        \includegraphics[width=\columnwidth]{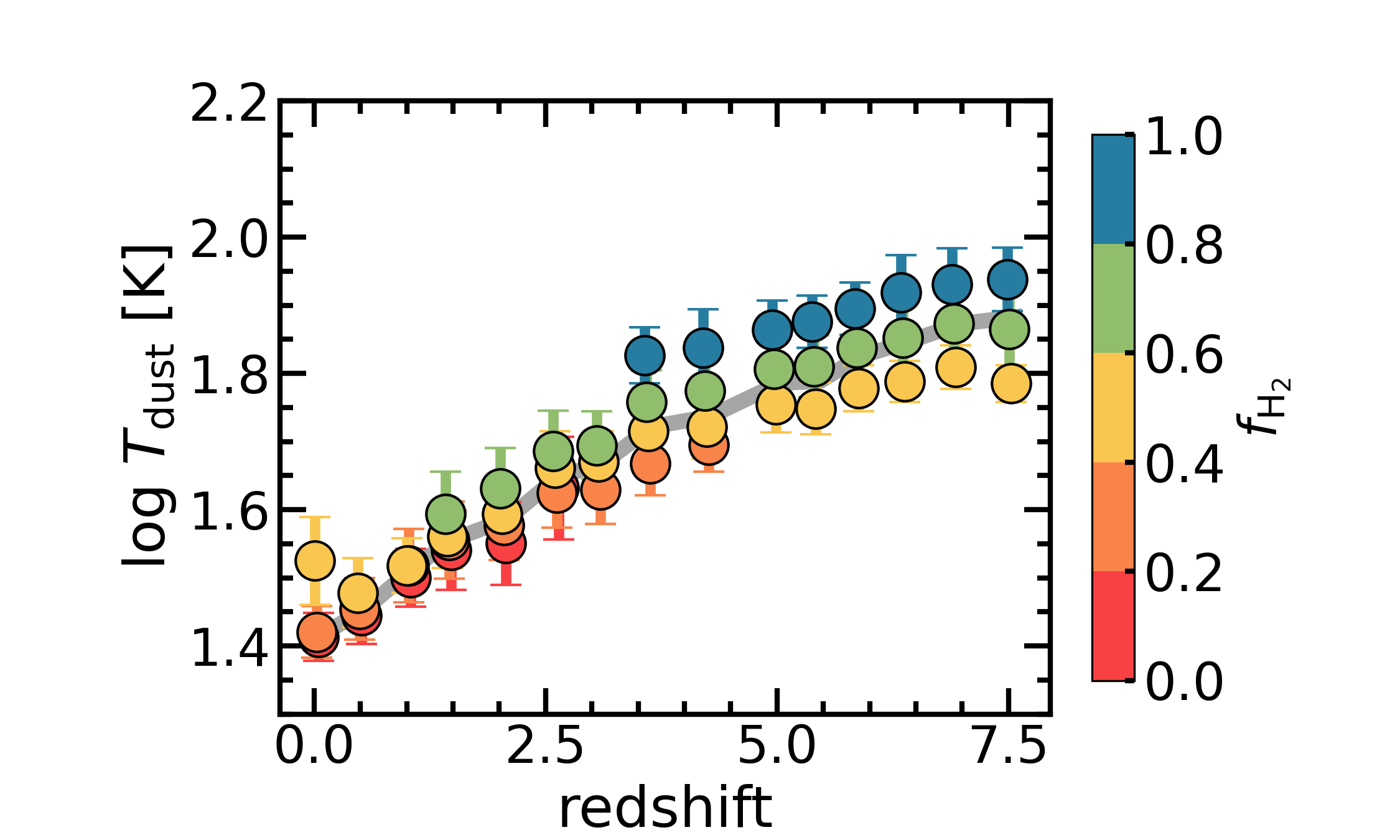}
    \end{subfigure}

    \vskip -0.4cm

    \begin{subfigure}[b]{\columnwidth}
        \centering
        \includegraphics[width=\columnwidth]{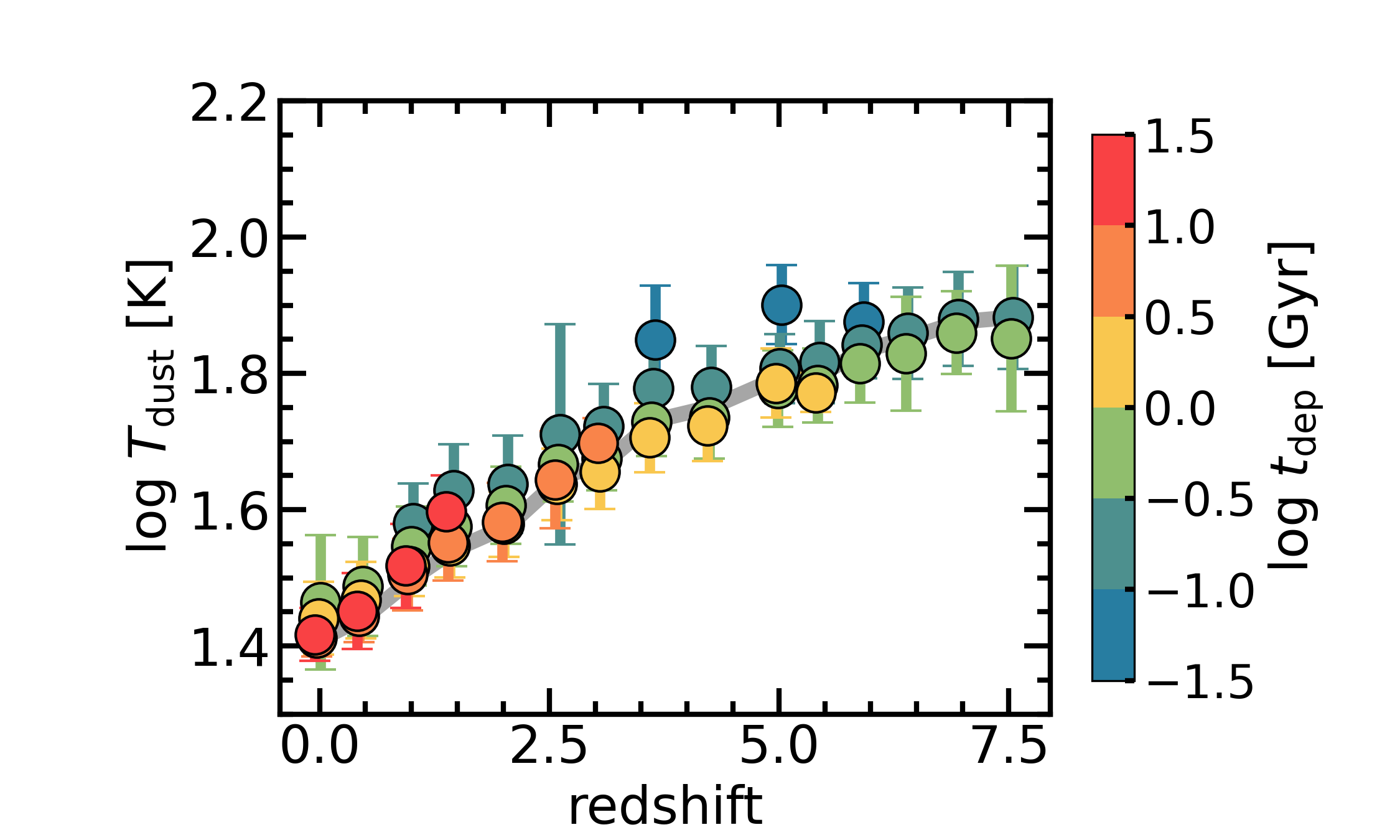}
    \end{subfigure}

    \caption{Dust temperature evolution with redshift separating galaxies in bins of specific SFR (top panel), molecular fraction (mid panel) and depletion time (bottom panel). Each panel shows the median and $16-84$th percentiles of the distribution. Colors indicate bins in the corresponding property. The gray line in each panel refers to the full population.}
    \label{fig:Tdust_binnings}
\end{figure}

\begin{figure}[t]
    \centering

    \begin{subfigure}[b]{\columnwidth}
        \centering
        \includegraphics[width=\columnwidth]{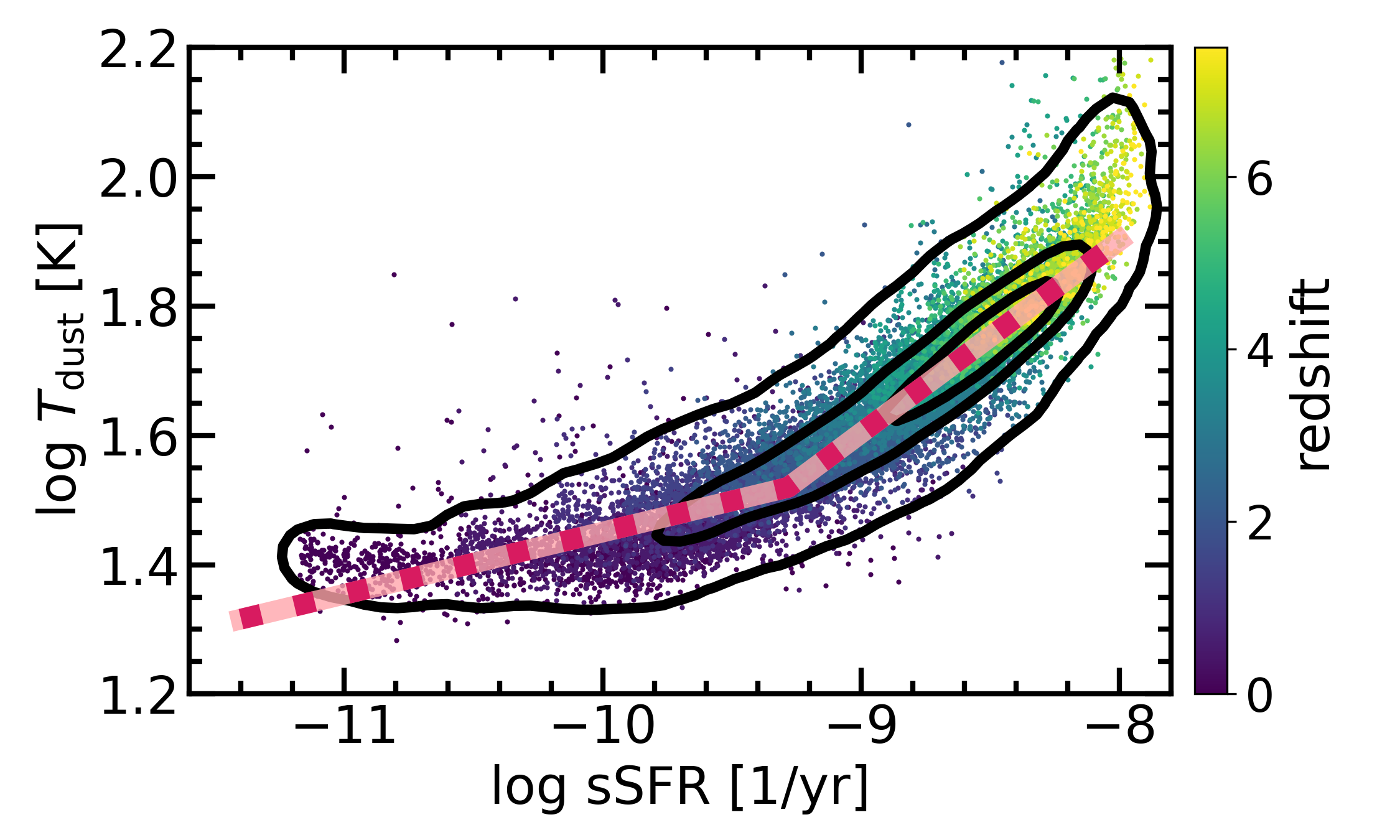}
    \end{subfigure}

    \begin{subfigure}[b]{\columnwidth}
        \centering
        \includegraphics[width=\columnwidth]{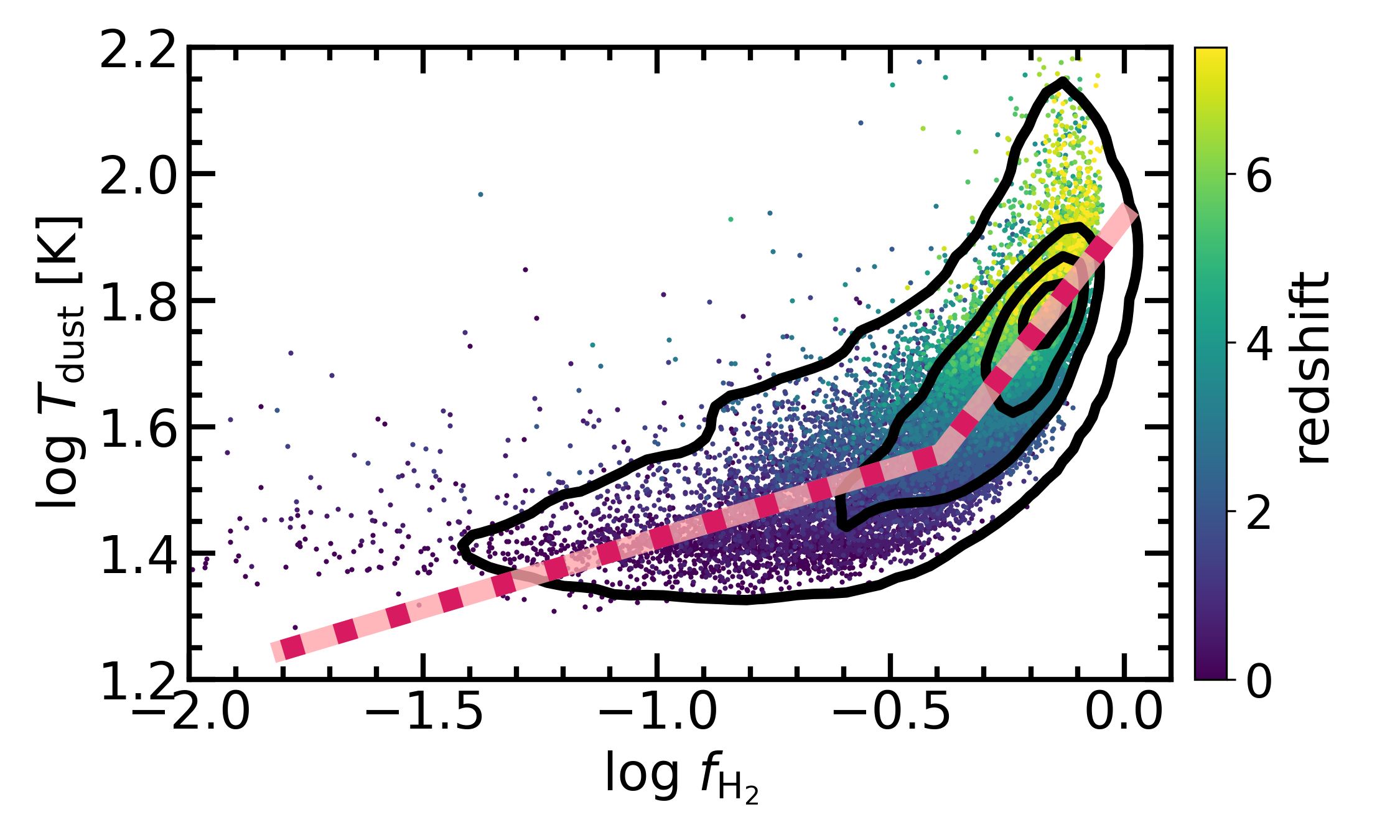}
    \end{subfigure}

    \begin{subfigure}[b]{\columnwidth}
        \centering
        \includegraphics[width=\columnwidth]{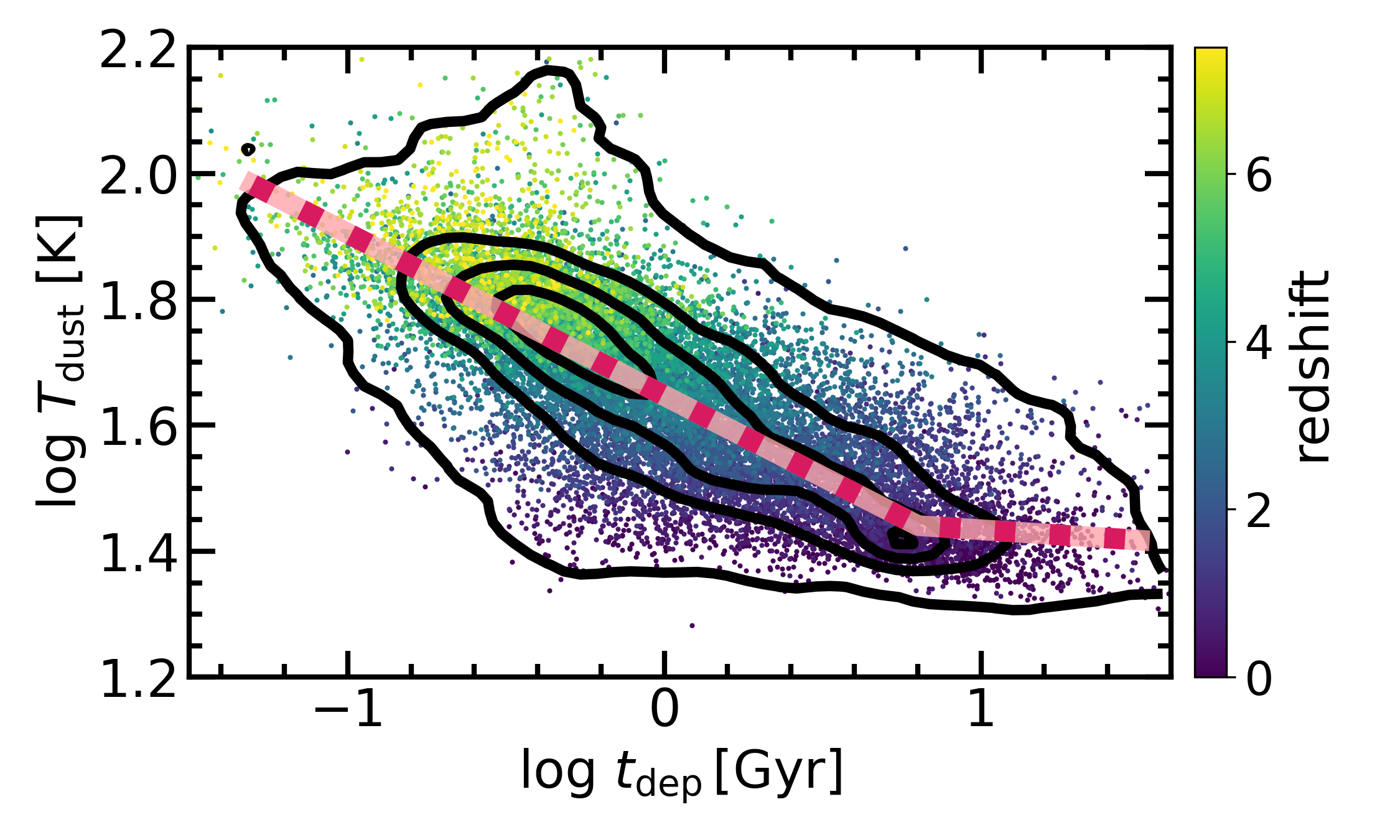}
    \end{subfigure}

    \caption{Dust temperature as a function of sSFR, molecular fraction $f_{\rm H_2}$, and depletion time $t_{\rm dep}$. Each point represents a galaxy and is color coded by redshift, while the black contours correspond to the full sample. The red dashed line shows a broken power law fit to data (Tab. \ref{tab:coefffit}).}
    \label{fig:Tdust_scaling2}
\end{figure}

This appendix presents the same analysis shown in the main text for the two most influential properties affecting dust temperature ($\Sigma_{\rm SFR}$ and DTG) identified by the Shapley analysis (Sect. \ref{sec:shap}), but now applied to the other three key properties: specific SFR, molecular gas fraction, and depletion time (see Tab. \ref{tab:FI}).\\
Fig. \ref{fig:Tdust_binnings} shows the redshift evolution of \tdust as predicted by our model, binned according to these galaxy properties. Fig. \ref{fig:Tdust_scaling2} illustrates the relations between dust temperature and these properties, along with their best-fitting broken power-laws (see Tab. \ref{tab:coefffit}).


\end{document}